\documentclass[english]{article}
\usepackage{lmodern}
\usepackage[T1]{fontenc}
\usepackage[latin9]{inputenc}
\usepackage{geometry}
\usepackage{babel}
\usepackage{amsmath}
\usepackage{amssymb}
\usepackage{graphicx}
\usepackage{setspace}
\usepackage{wasysym}
\usepackage[unicode=true,pdfusetitle,
 bookmarks=true,bookmarksnumbered=false,bookmarksopen=false,
 breaklinks=false,pdfborder={0 0 1},backref=false,colorlinks=false]
 {hyperref}
\hypersetup{
 colorlinks,linkcolor=blue,citecolor=blue,urlcolor=blue}

\makeatletter

\providecommand{\tabularnewline}{\\}

\newcommand{\lyxaddress}[1]{
	\par {\raggedright #1
	\vspace{1.4em}
	\noindent\par}
}

\date{}

\makeatother

\begin{document}
\title{Theory of spin--charge-coupled transport in proximitized graphene:
An SO(5) algebraic approach}
\author{Aires Ferreira$^{\dagger}$}
\maketitle

\lyxaddress{\begin{center}
$^{\dagger}$Department of Physics and York Centre for Quantum Technologies,\linebreak{}
University of York, YO10 5DD, York, United Kingdom
\par\end{center}}
\begin{abstract}
Establishing the conditions under which orbital, spin and lattice-pseudospin
degrees of freedom are mutually coupled in realistic nonequilibrium
conditions is a major goal in the emergent field of graphene spintronics.
Here, we use linear-response theory to obtain a unified microscopic
description of spin dynamics and coupled spin--charge transport in
graphene with an interface-induced Bychkov--Rashba effect. Our method
makes use of an SO(5) extension of the familiar inverse-diffuson approach
to obtain a quantum kinetic equation for the single-particle density
matrix that treats spin and pseudospin on equal footing and is valid
for arbitrary external perturbations. As an application of the formalism,
we derive a complete set of drift--diffusion equations for proximitized
graphene with scalar impurities in the presence of electric and spin-injection
fields which vary slowly in space and time. Our approach is amenable
to a wide variety of generalizations, including the study of coupled
spin--charge dynamics in layered materials with strong spin--valley
coupling and spin--orbit torques in van der Waals heterostructures.
\pagebreak{}
\end{abstract}

\section{Introduction}

There is a current fundamental and technological interest in the harnessing
of spin--orbit-coupling (SOC) effects in nonmagnetic media, particularly
for the interconversion of charge and spin currents and the generation
of nonequilibrium spin polarization \cite{RevModPhys.91.035004,RevModPhys.89.025006}.
A recent trend is the use of two-dimensional (2D) materials to engender
electrical control over SOC effects benefiting from their reduced
dimensionality and unique opto-electronic properties afforded by atomically
thin crystals and their heterostructures  \cite{RevModPhys.92.021003,Sierra_21}.
With graphene well established as a high-fidelity spin channel material
supporting room-temperature spin transport over lengths up to tens
of micrometers \cite{Tombros_2007,Han_11,Zomer_12,Kamalakar_15,Yan_16,Gebeyehu_2019},
an important challenge concerns the manipulation of nonequilibrium
spins by pure electrical means for future spin-logic applications.
While at first glance this might seem hopeless given the absence of
bulk ferromagnetism in graphene \cite{Sepioni_10}, not to mention
its ultra-low intrinsic SOC \cite{SOC_Graphene_Sichau_19}, several
strategies have been proposed to overcome this bottle neck, including
adatom engineering \cite{Ferreira_14,Balakrishnan_14} and proximity
effects achieved via van der Waals coupling to high-SOC 2D materials
\cite{Avsar_14,Wang_14,Wang_16,Gmitra_16,Island_19}. The latter approach
has shown great promise because the proximity-induced SOC can be well
resolved in energy (i.e., on the order of the quasiparticle broadening),
which facilitates the experimental demonstration of SOC effects with
reproducibility, e.g. by means of low-field magnetotransport measurements
\cite{Yang_2016,Yang_17,Volkl_17,Zihlmann_18,Wakamura_18}. Furthermore,
the strong interplay of spin and lattice-pseudospin degrees of freedom
in honeycomb layered materials gives rise to fingerprints of unique
hallmarks of SOC in  spin transport experiments. Most notably, the
emergence of spin--helical 2D Dirac fermions in van der Waals heterostructures
due to the Bychkov--Rashba (BR) effect \cite{BR_Graphene_Rashba_09}
has been predicted \cite{BR_Graphene_Offidani_17} and demonstrated
experimentally \cite{Ghiasi_19,Benitez_20,Li_2020,Hoque_21} to enable
efficient spin--charge interconversion at room temperature. Owing
to a unique spin--pseudospin entanglement of electronic wavefunctions,
the sign and magnitude of the nonequilibrium spin polarization in
BR-coupled graphene can be tuned with a back-gate voltage \cite{BR_Graphene_Offidani_17},
in contrast to spin-galvanic effects generated by topological insulators
\cite{Kondou_16}. Another interesting manifestation of proximity-induced
SOC is observed in Hanle-type spin precession experiments \cite{Ghiasi2017,Benitez2017},
where a rather unconventional spin dynamics results in spins polarized
in the plane of graphene relaxing about ten times faster than out-of-plane
spins. This effect, originally predicted by Cummings and co-workers
\cite{Cummings2017}, is explained by the combined action of interface-induced
spin--valley coupling and intervalley scattering triggered by point
defects, which opens an additional Dyakonov-Perel-like relaxation
channel for in-plane spins.

\bigskip{}

Due to the fast experimental progress in the field, the theory is
generally lagging but there are notable exceptions. A microscopic
theory of spin dynamics for graphene-based van der Waals heterostructures
with $C_{3v}$ point-group symmetry was put forward in Ref. \cite{Manuel2018}.
Moreover, a controlled diagrammatic approach to calculating linear
response functions in the presence of disorder and generic proximity
effects was developed in a recent series of works \cite{BR_Graphene_Offidani_17,BR_Graphene_Milletari_17,BR_Graphene_Offidani_MDPI_18,Offidani_Ferreira_18,Sousa_2020}.
An electrical detection scheme that enables full disentanglement of
competing SOC transport effects in diffusive lateral spin-valve devices
was also proposed recently in Ref. \cite{Cavill_20}. These early
works highlighted the key role played by symmetry, quantum geometry
and impurity scattering in the spin dynamics and spin--orbit-coupled
transport phenomena, such as the spin Hall effect (SHE) \cite{Hirsch_99},
in honeycomb layers. The diagrammatic approach has proven ideal for
the study of the weak-disorder limit relevant for clean samples with
large mean free paths, where numerical simulations have traditionally
struggled \cite{Garcia_18,Joao_20}. Interestingly, the emergence
of noncoplanar $\mathbf{k}$-space spin textures in broken inversion
symmetry conditions was shown to allow for a robust skew-scattering-induced
SHE that dominates over the intrinsic (spin-Berry-curvature) contribution
in the clean limit, while not requiring spin-dependent disorder potentials
\cite{BR_Graphene_Milletari_17}. On the other hand, tight-binding
methods have provided useful insights into the highly disordered limit
(e.g. via simulations of extrinsic SHE efficiency in samples with
a high coverage of adatoms \cite{Tuan_16}). However, a unified semiclassical
description of spin--orbit-coupled transport effects in the presence
of generic time- and space-dependent perturbations is still lacking,
even for the simplest case of $C_{6v}$-symmetric graphene heterostructures.
The aim of this paper is to fill this gap by developing a theoretical
framework that encompasses all the known phenomenology and has the
potential to provide new predictions for the numerous opto-spintronic
phenomena supported by van der Waals materials \cite{Cavill_20,Garcia_18,review_valleytronics_16,Xiao_12,Luo_17,Avsar_17}.
To that end, we devise a Green's function \emph{inverse diffuson}
approach that respects the SO(5) algebraic structure of spin--orbit-coupled
2D Dirac fermions (rather than projecting out the sublattice degree
of freedom to obtain a simplified description in terms of Bloch-type
equations \cite{Wang_11}), which can be used to derive quantum kinetic
equations in a simple and mathematically transparent fashion. This
extension of the powerful \emph{diffuson} approach originally developed
for 2D electron gases \cite{2DEG_SOC_Burkov_04,Wenk_10} will allow
us not only to keep track of the entangled dynamics of pseudospin
and spin observables, but also to obtain useful analytical expressions
for linear response functions to generalized fields of experimental
relevance, such as Zeeman-type spin-injection fields. 

This paper is organized as follows. In Sec. \ref{subsec:1.1}, we
introduce the 2D Dirac-Rashba model capturing the low-energy dynamics
of $C_{6v}$-symmetric graphene heterostructures. Sec. \ref{subsec:1.2}
describes the diagrammatic framework and the semiclassical approximation
used in treating scattering effects. In Sec. \ref{sec:2-Results}
we derive the semiclassical drift--diffusion transport equations
for the experimentally accessible macroscopic observables and discuss
various features and applications of the formalism. Sec. \ref{sec:Summary-and-outlook}
presents our conclusions.

\subsection{The 2D Dirac-Rashba model \label{subsec:1.1}}

\begin{figure}
\centering{}\centering{}\includegraphics[width=0.9\columnwidth]{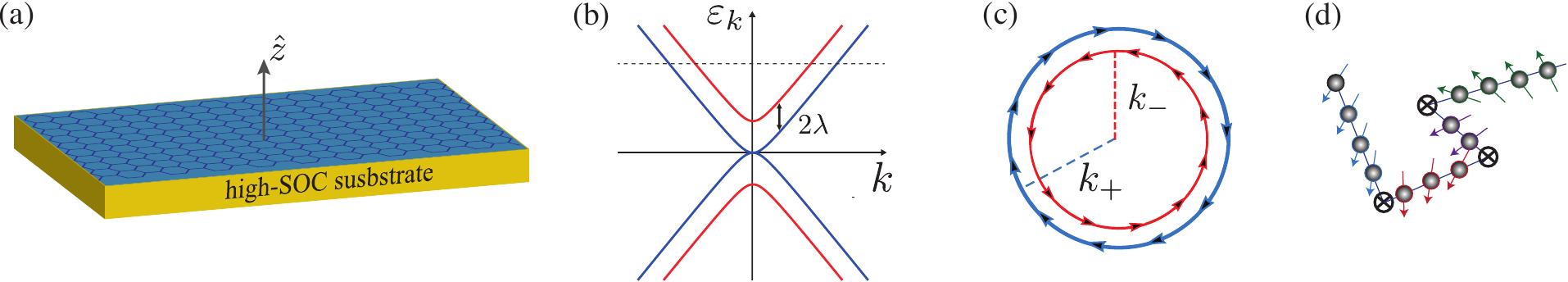}\caption{\label{fig:schematic}(a) Graphene placed on a substrate. The breaking
of mirror reflection symmetry $z\rightarrow-z$ is responsible for
the emergence of a BR effect. (b) Energy dispersion near a $K$ point.
The BR coupling lifts the spin degeneracy and opens a spin gap. Blue
(red) curves correspond to spin majority (minority) bands. (c) Tangential
winding of the spin texture outside the spin gap ($|\varepsilon|>2|\lambda|$).
(d) Dyakonov-Perel spin-relaxation mechanism is operative in samples
with a weak BR effect ($|\lambda|\ll\eta$, with $\eta$ the quasiparticle
broadening). Crosses indicate scattering centers and colored arrows
the spin vector.}
\end{figure}

For our theoretical study of spin--charge-coupled transport in spin--orbit-coupled
graphene, we use a model of noninteracting electrons subject to a
random impurity potential. Because we are interested in emergent phenomena
stemming from interfacial breaking of inversion symmetry, we assume
that the BR interaction is the dominant SOC effect (see Figs. \ref{fig:schematic}(a)
and (b)) \cite{BR_Graphene_Rashba_09,BR_Bychkov_Rashba_84}. The effective
Hamiltonian at the $K$ valley can be written in terms of tensor products
of Pauli matrices $\sigma_{a}\otimes s_{b}$ ($a,b=x,y,z$) acting
on the pseudospin--spin space as \cite{note:1}
\begin{equation}
H=\int d^{2}\mathbf{x}\,\Psi^{\dagger}(\mathbf{x})\,\left[\,v\sigma_{\mu}\left(p^{\mu}+\boldsymbol{\mathcal{A}}^{\mu}\right)+V(\mathbf{x})\,\right]\Psi(\mathbf{x})\,,\label{eq:Hamiltonian}
\end{equation}
where $v\simeq10^{6}$ m/s is the bare Fermi velocity of the massless
Dirac fermions, $p^{\mu}=(-\varepsilon/v,-\imath\hbar\nabla)$ is
the energy--momentum 3-vector, $\boldsymbol{\mathcal{A}}^{\mu}=\sum_{a=x,y,z}\mathcal{A}_{a}^{\mu}\,s_{a}$
($\mu=0,x,y,z)$ is the non-Abelian SU(2) gauge field capturing all
spin-dependent effects and $V(\mathbf{x})$ is the impurity potential.
Here, summation over repeated indices is implied with $\sigma_{0}$
denoting the identity matrix. The BR effect with coupling strength
$\lambda$ is encoded in the gauge field components $\mathcal{A}_{y}^{x}=-\mathcal{A}_{x}^{y}=\lambda/v$,
which generate the corresponding spin--orbit interaction via minimal
coupling {[}$\mathcal{H}_{\textrm{BR}}=\lambda\,(\boldsymbol{\sigma}\times\boldsymbol{s})_{z}${]}.
Equation (\ref{eq:Hamiltonian}) has been dubbed the 2D Dirac--Rashba
model in recent literature \cite{BR_Graphene_Offidani_17,BR_Graphene_Milletari_17}.
We are not concerned with purely extrinsic transport phenomena induced
by spin--orbit-active impurities (i.e. random SOC), which have been
the object of detailed microscopic analysis in early work \cite{Ferreira_14,Pachoud_14,Yang_16,Huang_16,SOC_Graphene_Milletari_Ferreira16a,Milletari_16_b}.
Rather, our focus here is on the transport properties of spin--helical
2D Dirac fermions realized in graphene-based heterostructures with
a dilute concentration of spin-transparent impurities and a well-established
(i.e. spatially uniform) BR effect.

The dispersion relation of the minimal Dirac--Rashba model, $\mathcal{H}_{0\mathbf{k}}=\hbar v\,\boldsymbol{\sigma}\cdot\mathbf{k}+\mathcal{H}_{\textrm{BR}}$,
reads as
\begin{equation}
\varepsilon_{\mu\nu}(\mathbf{k})=\mu\lambda+\nu\sqrt{\lambda^{2}+\hbar^{2}v^{2}k^{2}}\,,\label{eq:spectrum}
\end{equation}
where $\mu(\nu)=\pm1$ labels, respectively, the spin-helicity and
polarity of charge carriers (Fig.\,\ref{fig:schematic}(b)). The
BR effect lifts the spin degeneracy and locks the spin polarization
and wavevector at right angles, leading to a spin-helical configuration
in $\mathbf{k}$-space (Fig.\,\ref{fig:schematic}(c)). The Dirac
nature of the carriers enables a low electronic density regime characterized
by a simply-connected Fermi surface with well defined spin-helicity
(i.e. $\mu=-1$), akin to the surface states of topological insulators,
with interesting consequences for spin-charge conversion effects \cite{Manuel2018}.

The equilibrium spin polarization associated with the Bloch eigenstates
of $\mathcal{H}_{0\mathbf{k}}$ is easily computed as
\begin{equation}
\langle\mathbf{S}\rangle_{\mu\nu\mathbf{k}}=\frac{\hbar}{2}\langle\boldsymbol{s}\rangle_{\mu\nu\mathbf{k}}=-\frac{\hbar}{2}\mu\,\langle\boldsymbol{\sigma}\rangle_{\mu\nu\mathbf{k}}\times\mathbf{\hat{z}}\,,\label{eq:SP}
\end{equation}
where $\langle\boldsymbol{\sigma}\rangle_{\mu\nu\mathbf{k}}=(1/\hbar v)\nabla_{\mathbf{k}}\epsilon_{\mu\nu}(\mathbf{k})$
is the expectation value of the pseudospin polarization vector. The
spin winding of the Fermi surface is shared with other surfaces possessing
broken inversion symmetry, but unlike conventional spin-helical states,
its energy dependence reflects a spin-angular momentum transfer between
spin and pseudospin channels. Such a spin-pseudospin coupling is responsible
for the conspicuous wavevector dependence of the equilibrium spin
texture {[}Eq.~(\ref{eq:SP}){]}. Indeed, the spin-polarization magnitude
is vanishing at the corners of the first Brillouin zone, where the
band velocity is vanishing ($\langle\boldsymbol{\sigma}\rangle_{\mu\nu\mathbf{k}\rightarrow0}=0$).
Away from the zone corners $K$ and $K^{\prime}$, the spin texture
magnitude increases monotonically with the Fermi wavevector until
it saturates away from the spin gap (i.e. for $|\varepsilon|\apprge2|\lambda|$).
As noted by Rashba \cite{BR_Graphene_Rashba_09}, the strong momentum-dependence
of the spin texture at low energies is a unique fingerprint of BR-coupled
2D Dirac fermions. 

The random potential in Eq. (\ref{eq:Hamiltonian}), which in this
work is assumed to be scalar, affects the spin dynamics of charge
carriers by inducing scattering between electronic states with different
effective Larmor fields $\boldsymbol{\Omega}_{\mu\nu\mathbf{k}}=\lambda\langle\mathbf{s}\rangle_{\mu\nu\mathbf{k}}\approx-\mu\nu\lambda\,\hat{\mathbf{k}}\times\hat{z}$
for $|\varepsilon|\gg|\lambda|$. Due to the random change of precession
axis, an initial nonequilibrium spin polarization will decay exponentially
with time. In the standard weak-SOC regime (realized in systems with
a small spin splitting compared to the disorder-induced broadening
\cite{RevModPhys.76.323,Wu_review,Boross_13}), the spin-relaxation
rate is $\tau_{s}^{-1}\propto\lambda^{2}\tau$, where $\tau$ is the
elastic scattering time (Fig. \ref{fig:schematic}(d)). Interestingly,
the spin-relaxation rate of out-of-plane spins (i.e. polarized along
the $\hat{z}$-axis) is twice that of in-plane spins ($\tau_{s,\perp}/\tau_{s,\parallel}=1/2$),
which could provide an experimentally detectable signature of BR effect
using Hanle-type spin precession measurements \cite{Raes_16,Ringer_18}.
Impurity scattering also plays a key role in coupled spin--charge
transport phenomena, profoundly affecting the efficiency of spin Hall
and spin-charge conversion (spin-galvanic) effects, even in the clean
limit with $|\varepsilon|\tau\gg1$, as we shall see below. 

\subsection{Theoretical framework\label{subsec:1.2}}

To derive a rigorous microscopic picture of coupled spin-charge transport,
we evaluate the density matrix response function employing many-body
perturbation theory methods \cite{Book_Fetter,Doniach,Mahan}. Our
aim is to generalize the familiar \emph{diffuson} approach for 2D
electron gases \cite{2DEG_SOC_Burkov_04,Wenk_10} to accommodate the
enlarged SO$(5)$ Clifford algebra of 2D Dirac fermions. Because we
are interested in the diffusive regime realized in weakly disordered
samples with $|\varepsilon|\tau\gg1$, we neglect quantum corrections
arising from weak localization and higher-order spin-orbit scattering
effects, such as quantum side jumps and diffractive skew scattering
\cite{SOC_Graphene_Milletari_Ferreira16a,Milletari_16_b,Ado_2015}.
In the diagrammatic language, such semiclassical approximation amounts
to discarding crossing diagrams encoding coherent multiple impurity
scattering events, as well as higher-order noncrossing diagrams with
a perturbation parameter $1/(|\varepsilon|\tau)\ll1$ \cite{SOC_Graphene_Milletari_Ferreira16a}.
Unless stated otherwise, we work in natural units with $\hbar\equiv1\equiv e$.
Additionaly, for ease of notation, we assume throughout that $\varepsilon,\lambda>0$. 

The central object in our approach is the real-time retarded(R)/advanced(A)
single-particle Green's function ($a=A,R\equiv-,+$) defined as

\begin{equation}
G^{a}(\mathbf{x},\mathbf{x}^{\prime};t-t^{\prime})=\mp\imath\left\langle 0|T\left[\Psi(\mathbf{x},t),\Psi^{\dagger}(\mathbf{x}^{\prime},t^{\prime})\right]|0\right\rangle \theta(\pm t\mp t^{\prime}),\label{eq:time-ordered-G}
\end{equation}
where $\Psi^{\dagger}(\mathbf{x},t)$ {[}$\Psi(\mathbf{x},t)${]}
are 4-component field operators creating (annihilating) a 2D Dirac
fermion at position $\mathbf{x}$ and time $t$, $T$ is the time-ordering
symbol and $\theta(.)$ is the Heaviside step function. After disorder
averaging (indicated below by an overline), the Green's function in
momentum-frequency space acquires the familiar form
\begin{equation}
\mathcal{G}_{\mathbf{k}}^{a}(\varepsilon)=\frac{1}{[\mathcal{G}_{0\mathbf{k}}^{a}(\varepsilon)]^{-1}-\Sigma_{\mathbf{k}}^{a}(\varepsilon)}\,,\label{eq:av_Greens_Function_}
\end{equation}
where $\mathcal{G}_{0\mathbf{k}}^{a}(\varepsilon)=[\varepsilon-v\boldsymbol{\sigma}\cdot\mathbf{k}-\lambda\,\left(\boldsymbol{\sigma}\times\mathbf{s}\right)\cdot\hat{z}+\imath\,a\,0^{+}]^{-1}$
is the Fourier transform of the clean Green's function and
\begin{equation}
\Sigma_{\mathbf{k}}^{a}(\varepsilon)=\int d\mathbf{(x-x^{\prime})}\,e^{-\imath\mathbf{k}(\mathbf{x}-\mathbf{x}^{\prime})}\overline{\langle\mathbf{x}^{\prime}|V\frac{1}{1-G_{0}^{a}(\varepsilon)V}|\mathbf{x}\rangle}\,\label{eq:self-energy_NC}
\end{equation}
is the quasiparticle self energy in the noncrossing approximation.
For uncorrelated short-range impurity potentials, the self energy
is $\mathbf{k}$-independent, and so we set $\Sigma_{\mathbf{k}}^{a}(\varepsilon)\equiv\Sigma^{a}(\varepsilon)$
from here onwards. 

\begin{figure}
\begin{centering}
\includegraphics[width=0.9\columnwidth]{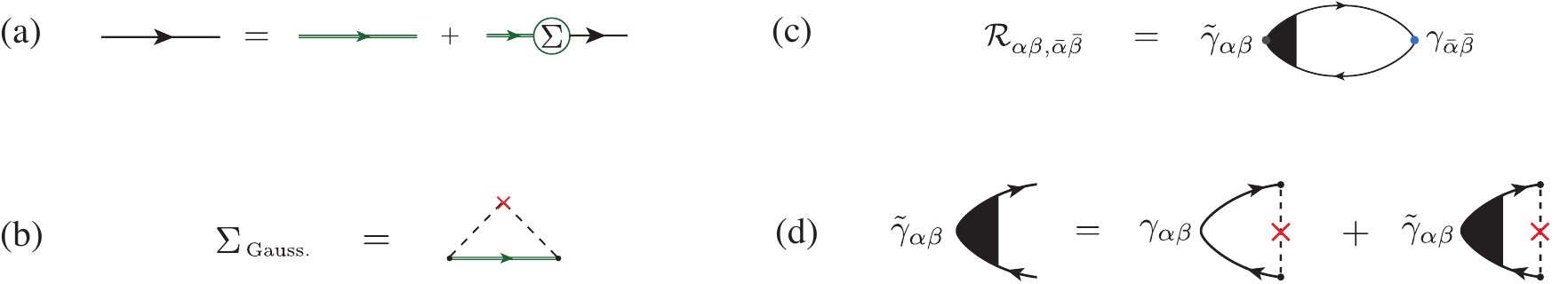}
\par\end{centering}
\caption{\label{fig:SO5_diagrammatic_scheme}(a-d) Diagrammatic scheme for
the evaluation of the linear response density matrix. Green (black)
solid line with an arrow denotes the free (disorder-averaged) Green's
function. Dashed lines depict scattering potential insertions ($u_{0}$)
and the cross represents the impurity density ($n_{i}$).}
\end{figure}

In order to keep our discussion as simple as possible, we compute
the self energy under the assumption of weak Gaussian disorder. This
will simplify our analytical treatment significantly while capturing
the essential physics \cite{note:2}. Diagrammatically, the weak disorder
approximation amounts to retaining only the contribution from the
'rainbow' diagram with two impurity potential lines; see Figs. \ref{fig:SO5_diagrammatic_scheme}
(a)-(b). Replacing the explicit form of the two-point correlator $\overline{V(\mathbf{x})V(\mathbf{x}^{\prime})}=n_{i}u_{0}^{2}\,\delta(\mathbf{x}-\mathbf{x}^{\prime})$
in the Dyson expansion of the self energy, where $n_{i}$ is the impurity
areal density and $u_{0}$ is the potential strength, one arrives
at
\begin{equation}
\Sigma^{\pm}(\varepsilon)=\mp\,\left(\imath\,/2\tau\right)\left[\theta(\varepsilon-2\lambda)+\theta(2\lambda-\varepsilon)\left(1+\sigma_{z}s_{z}-\frac{\varepsilon}{\lambda}\gamma_{r}\right)\frac{\lambda}{\varepsilon}\right]\,,\label{eq:self_energy}
\end{equation}
where $1/2\tau\equiv\eta=n_{i}\,u_{0}^{2}\varepsilon/4v^{2}$ is the
quasiparticle broadening and $\gamma_{r}\equiv(\sigma_{x}s_{y}-\sigma_{y}s_{x})/2$.
The existence of two distinct transport regimes at low energies (i.e.
$\varepsilon>2\lambda$ and $\varepsilon<2\lambda$) is a unique feature
of the 2D Dirac--Rashba model, which is responsible for the existence
of a maximum in current-induced spin polarization efficiency when
the (gate-tunable) Fermi energy lies precisely at the spin-gap edge
\cite{BR_Graphene_Offidani_17}.\textbf{ }

In this paper, we are primarily interested in the diffusive coupled
spin--charge dynamics which occur in graphene flakes with weak proximity-induced
SOC at moderate--high charge carrier densities ($\varepsilon\tau\gg1\gg\lambda\tau$)
\cite{BR_Graphene_Offidani_MDPI_18}. The condition $\varepsilon\gg\lambda$
implies that the BR-slit bands with opposite spin helicities are occupied
at the Fermi level. To study the behavior of the system away from
equilibrium, it is convenient to introduce the generalized one-particle
density operator 
\begin{equation}
\hat{\rho}_{\alpha\beta}(\mathbf{x};t)=\Psi^{\dagger}(\mathbf{\mathbf{x}},t)\,\mathcal{\gamma}_{\alpha\beta}\,\Psi(\mathbf{x},t)\,,\label{eq:density_op}
\end{equation}
 where $\mathcal{\mathcal{\gamma}}_{\alpha\beta}=\sigma_{\alpha}\otimes s_{\beta}$
(with $\alpha,\beta=0,x,y,z$) span the vector space of Hermitian
$4\times4$ matrices. The density matrix characterizing a given nonequilibrium
state can be expanded as a linear combination of the Clifford algebra
basis elements, such that
\begin{equation}
\rho(\mathbf{x},t)=\frac{1}{d}\sum_{\alpha,\beta=0,x,y,z}\gamma_{\alpha\beta}\,\langle\hat{\rho}_{\alpha\beta}(\mathbf{x},t)\rangle,\label{eq:density_matrix}
\end{equation}
where $\langle...\rangle$ denotes quantum and disorder averages and
$d=\textrm{dim}\,H\equiv4$ is a normalization factor. The expectation
value of a generic local observable, $\mathcal{O}=\sum_{\alpha\beta}\mathcal{O}_{\alpha\beta}\gamma_{\alpha\beta}$,
is obtained according to
\begin{equation}
\langle\mathcal{O}(\mathbf{x},t)\rangle:=\text{tr}\left[\mathcal{O}\,\rho(\mathbf{x},t)\right]=\sum_{\alpha,\beta}\mathcal{O}_{\alpha\beta}\,\rho_{\alpha\beta}(\mathbf{x},t)\,,\label{eq:exp}
\end{equation}
where $\textrm{tr}$ indicates the trace over internal degrees of
freedom and $\rho_{\alpha\beta}(\mathbf{x},t)\equiv\langle\hat{\rho}_{\alpha\beta}(\mathbf{x},t)\rangle$. 

In this work we are concerned with the semiclassical dynamics of typical
spin transport observables, such as the spin-polarization density
and the spin current density. Thus, it is more convenient to work
directly with the deviation from equilibrium of the expectation values,
i.e. $\langle\delta\mathcal{O}(\mathbf{x},t)\rangle:=\text{tr}\left[\mathcal{O}\,\delta\rho(\mathbf{x},t)\right]$
with $\delta\rho_{\alpha\beta}(\mathbf{x},t):=\rho_{\alpha\beta}(\mathbf{x},t)-\rho_{\alpha\beta}^{0}$.
Here, $\rho_{\alpha\beta}^{0}$ denotes the equilibrium part of the
(disorder-averaged) density matrix. The macroscopic observables of
interest to us are the charge density, $N$, spin polarization density,
$S^{a}$ ($a=x,y,z)$, charge current density, $J_{i}$ ($i=x,y$),
and spin current density, $\mathcal{J}_{i}^{a}$. The corresponding
expectation values away from equilibrium are defined as
\begin{equation}
\{\delta N,\delta S^{a},\delta J_{i},\delta\mathcal{J}_{i}^{a}\}=\{-e\delta\rho_{00},\,\frac{\hbar}{2}\delta\rho_{0a},\,-ev\delta\rho_{i0},\,\frac{\hbar v}{2}\delta\rho_{ia}\}\,,\label{eq:local_observables_definition}
\end{equation}
 where we have temporarily reinstated $\hbar$ and $e$ to distinguish
between charge and spin currents. 

The interaction Hamiltonian is $V(t)=\int d^{2}\mathbf{x}\,\Psi^{\dagger}(\mathbf{x})\,\left[\mathcal{H}_{\textrm{int}}(\mathbf{x},t)\right]\Psi(\mathbf{x})$
with 
\begin{equation}
\mathcal{H}_{\textrm{int}}(\mathbf{x},t)=v\sum_{\alpha,\beta}\gamma_{\alpha\beta}\,\mathcal{A}_{\alpha\beta}^{\textrm{ext}}(\mathbf{x},t)\,,\label{eq:SO4_vec_pot}
\end{equation}
 where $\mathcal{A}_{\alpha\beta}^{\textrm{ext}}(\mathbf{x},t)$ ($\alpha,\beta=0,x,y,z$)
spans the Clifford algebra and thus describes any type of charge--spin
perturbation applied to the system. In Sec. \ref{subsec:2.diffuson_ham},
we shall show that the linear response density matrix $\delta\rho(\mathbf{x},t)$,
when properly coarse-grained over typical length and time scales (i.e.,
$|\mathbf{x}|\gg l\equiv v\tau$ and $t\gg\tau$), is governed by
an enlarged $16\times16$ \emph{diffuson Hamiltonian} that is the
SO(5) analogue of the familiar inverse density fluctuation propagator
of 2D electron gases \cite{2DEG_SOC_Burkov_04,Wenk_10,Shen_14} and
topological insulators \cite{Burkov_10,Sacksteder_12}. In Sec. \ref{subsec:2_unified-coarse_grained_dynamics},
the linear response machinery will be applied to derive the full set
of drift--diffusion transport equations for the variables $\delta N$,
$\delta S^{a}$, $\delta J_{i}$ and $\delta\mathcal{J}_{i}^{a}$,
and thus establish a rigorous microscopic picture for the coarse-grained
dynamics of the problem. In the following, we define $N\equiv\delta N$,
$S^{a}\equiv\delta S^{a}$, $J_{i}\equiv\delta J_{i}$ and $\mathcal{J}_{i}^{a}\equiv\delta\mathcal{J}_{i}^{a}$
for ease of notation.

\section{Results\label{sec:2-Results}}

\subsection{\emph{Diffuson} Dirac Hamiltonian\label{subsec:2.diffuson_ham}}

We start by setting up the formalism needed to derive a quantum kinetic
equation for BR-coupled 2D Dirac fermions. From standard linear response
theory, the zero temperature density matrix is given by 

\begin{equation}
\delta\rho_{\alpha\beta}(\mathbf{x},t)=\sum_{\bar{\alpha},\bar{\beta}=0,x,y,z}\int d\text{\textbf{x}}^{\prime}\int dt^{\prime}\:\mathcal{R}_{\alpha\beta,\bar{\alpha}\bar{\beta}}(\text{\textbf{x}}-\text{\textbf{x}}^{\prime},t-t^{\prime})\,v\mathcal{\mathcal{A}}_{\bar{\alpha}\bar{\beta}}^{\textrm{ext}}(\text{\textbf{x}}^{\prime},t^{\prime})\,,\label{eq:Kubo}
\end{equation}
where 
\begin{equation}
\mathcal{R}_{\alpha\beta,\bar{\alpha}\bar{\beta}}(\text{\textbf{x}}-\text{\textbf{x}}^{\prime},t-t^{\prime})=-\imath\theta(t-t^{\prime})\langle\overline{[\hat{\rho}_{\alpha\beta}(\mathbf{x},t),\hat{\rho}_{\bar{\alpha}\bar{\beta}}(\mathbf{x}^{\prime},t^{\prime})]}\rangle\,\label{eq:RRf}
\end{equation}
is the retarded response function. Equation (\ref{eq:RRf}) is best
evaluated in momentum--frequency space using the Green's-function
method \cite{Mahan}. A summation of noncrossing two-particle diagrams,
as depicted in Figs. \ref{fig:SO5_diagrammatic_scheme} (c)-(d), leads
to 
\begin{equation}
\mathcal{R}_{\alpha\beta,\gamma\delta}(\mathbf{q},\omega)\simeq\frac{\omega}{2\pi i}\,\sum_{\mathbf{k}}\,\text{tr}\left\{ \gamma_{\alpha\beta}\,\mathcal{G}_{\mathbf{k+q}}^{R}(\varepsilon+\omega)\,\tilde{\gamma}_{\gamma\delta}(\mathbf{q},\omega)\,\mathcal{G}_{\mathbf{k}}^{A}(\varepsilon)\right\} \,,\label{eq:density_matrix_response_function_R}
\end{equation}
  where the renormalized vertex operator $\tilde{\gamma}_{\alpha\beta}(\mathbf{q},\omega)$
satisfies the Bethe--Salpeter equation 
\begin{equation}
\tilde{\gamma}_{\alpha\beta}(\mathbf{q},\omega)=\gamma_{\alpha\beta}+n_{i}u_{0}^{2}\,\sum_{\text{\textbf{k}}}\,\mathcal{G}_{\mathbf{k+\mathbf{q}}}^{R}(\varepsilon+\omega)\,\tilde{\gamma}_{\alpha\beta}(\mathbf{q},\omega)\,\mathcal{G}_{\mathbf{k}}^{A}(\varepsilon)\,.\label{eq:BS_eq}
\end{equation}

Next, we project both sides of Eq.~(\ref{eq:BS_eq}) onto the Dirac
matrices and define
\begin{align}
\tilde{\boldsymbol{\gamma}}_{\alpha\beta} & (\mathbf{q},\omega)=(\tilde{\gamma}_{\alpha\beta00}(\mathbf{q},\omega),...,\tilde{\gamma}_{\alpha\beta zz}(\mathbf{q},\omega))^{T}\,,\label{eq:vertex_q_omega}
\end{align}
with $\tilde{\gamma}_{\alpha\beta\varrho\varsigma}(\mathbf{q},\omega)=(1/d)\,\textrm{tr}[\gamma_{\varrho\varsigma}\,\tilde{\gamma}_{\alpha\beta}(\mathbf{q},\omega)]$,
to obtain the renormalized vertex in a suitable closed form
\begin{equation}
\tilde{\mathbf{\boldsymbol{\gamma}}}_{\alpha\beta}(\mathbf{q},\omega)=(I-[M(\mathbf{q},\omega)]^{T})^{-1}\boldsymbol{\gamma}_{\alpha\beta}\,,\label{eq:vertex_closed_form}
\end{equation}
where $\boldsymbol{\gamma}_{\alpha\beta}=(0,...,1,...,0)^{T}$ is
an auxiliary vector with nonzero component $(\boldsymbol{\gamma}_{\alpha\beta})_{\alpha\beta}=1$,
$I$ is the identity matrix, $T$ denotes the transpose operation
and $M$ is a square matrix of ``bubbles'' with elements
\begin{equation}
M_{\mu\nu,\varrho\varsigma}(\mathbf{q},\omega)=\frac{n_{i}u_{0}^{2}}{d}\sum_{\text{\textbf{k}}}\text{tr}\left[\mathcal{G}_{\mathbf{k+\mathbf{q}}}^{R}(\varepsilon+\omega)\gamma_{\mu\nu}\mathcal{G}_{\mathbf{k}}^{A}(\varepsilon)\gamma_{\varrho\varsigma}\right]\,.\label{eq:bubbles}
\end{equation}

\begin{table}[b]
\begin{centering}
\begin{tabular}{|c|c|c|c|c|c|c|c|c|c|c|c|c|c|c|c|c|}
\cline{2-17} \cline{3-17} \cline{4-17} \cline{5-17} \cline{6-17} \cline{7-17} \cline{8-17} \cline{9-17} \cline{10-17} \cline{11-17} \cline{12-17} \cline{13-17} \cline{14-17} \cline{15-17} \cline{16-17} \cline{17-17} 
\multicolumn{1}{c|}{} & $\gamma_{00}$ & $\gamma_{01}$ & $\gamma_{02}$ & $\gamma_{03}$ & $\gamma_{10}$ & $\gamma_{11}$ & $\gamma_{12}$ & $\gamma_{13}$ & $\gamma_{20}$ & $\gamma_{21}$ & $\gamma_{22}$ & $\gamma_{23}$ & $\gamma_{30}$ & $\gamma_{31}$ & $\gamma_{32}$ & $\gamma_{33}$\tabularnewline
\hline 
$\mathcal{O}$ & $N$ & $S^{x}$ & $S^{y}$ & $S^{z}$ & $J_{x}$ & $\mathcal{J}_{x}^{x}$ & $\mathcal{J}_{x}^{y}$ & $\mathcal{J}_{x}^{z}$ & $J_{y}^ {}$ & $\mathcal{J}_{y}^{x}$ & $\mathcal{J}_{y}^{y}$ & $\mathcal{J}_{y}^{z}$ & $N_{s}$ & $S_{s}^{x}$ & $S_{s}^{y}$ & $S_{s}^{z}$\tabularnewline
\hline 
$C_{2}$ & +1 & $-1$ & $-1$ & $+1$ & $-1$ & $+1$ & $+1$ & $-1$ & $-1$ & $+1$ & $+1$ & $-1$ & $+1$ & $-1$ & $-1$ & $+1$\tabularnewline
\hline 
$R_{x}$ & +1 & $-1$ & $+1$ & $-1$ & $+1$ & $-1$ & $+1$ & $-1$ & $-1$ & $+1$ & $-1$ & $+1$ & $-1$ & $+1$ & $-1$ & $+1$\tabularnewline
\hline 
$\mathcal{T}$ & +1 & $-1$ & $-1$ & $-1$ & $-1$ & $+1$ & $+1$ & $+1$ & $-1$ & $+1$ & $+1$ & $+1$ & $-1$ & $+1$ & $+1$ & $+1$\tabularnewline
\hline 
\end{tabular}
\par\end{centering}
\centering{}\caption{\label{tab:Parity-of-Dirac}Classification of Dirac matrices and corresponding
observables under $C_{2}$-rotation, mirror-reflection $R_{x}$ and
time-reversal operation. Sublattice-staggered densities are indicated
with subscript \textquotedblleft$s$\textquotedblright .}
\end{table}
We may now recast the Fourier space response function into a more
compact form $\mathcal{R}_{\alpha\beta,\bar{\alpha}\bar{\beta}}(\mathbf{q},\omega)=-\imath\omega\nu_{0}\mathcal{D}_{\alpha\beta,\bar{\alpha}\bar{\beta}}(\mathbf{q},\omega)$,
where $\nu_{0}=\varepsilon/\pi v^{2}$ is the density of states per
spin and 
\begin{equation}
\mathcal{D}_{\alpha\beta,\bar{\alpha}\bar{\beta}}(\mathbf{q},\omega)=\tau\left(\tilde{\gamma}_{\bar{\alpha}\bar{\beta}\alpha\beta}(\mathbf{q},\omega)-\delta_{\alpha\bar{\alpha}}\delta_{\beta\bar{\beta}}\right)\,\label{eq:diffuson}
\end{equation}
 is the so-called \emph{diffuson }(here, $\delta_{\alpha\beta}$ is
the Kronecker delta symbol)\emph{.} Its inverse, the \emph{diffuson
Hamiltonian} $\mathcal{\mathcal{H}_{\mathcal{D}}}\equiv\mathcal{D}^{-1}$,
\begin{align}
\mathcal{\mathcal{H}_{\mathcal{D}}}(\mathbf{q},\omega) & =\frac{1}{\tau}\left(\left(I-[M(\mathbf{q},\omega)]^{T}\right)^{-1}-I\right)^{-1},\label{eq:diff_Ham_exp}
\end{align}
provides the kernel of the linear response quantum kinetic equation
\begin{equation}
\mathcal{H}_{\mathcal{D}}(\mathbf{q},\omega)\cdot\delta\vec{\rho}(\mathbf{q},\omega)=-\imath v\omega\nu_{0}\vec{\mathcal{A}}_{\textrm{ext}}(\mathbf{q},\omega)\,,\label{eq:diff_Ham_kinetic_Eq}
\end{equation}
where $\delta\vec{\rho}(\mathbf{q},\omega)\equiv(\delta\rho_{00}(\mathbf{q},\omega),...,,\delta\rho_{zz}(\mathbf{q},\omega))^{T}$
and $\vec{\mathcal{A}}_{\textrm{ext}}(\mathbf{q},\omega)\equiv(\mathcal{A}_{00}^{\textrm{ext}}(\mathbf{q},\omega),...,A_{zz}^{\textrm{ext}}(\mathbf{q},\omega))^{T}$
are the Fourier-space components of the one-particle density matrix
and generalized external vector potential, respectively.

Let us briefly discuss the general structure of $\mathcal{\mathcal{H}_{\mathcal{D}}}(\mathbf{q},\omega)$
in the long wavelength limit of interest to us. Zero entries in the
$16\times16$ bubble matrix $M(0,\omega)$ can be readily identified
by applying the following $C_{6v}$ point-group operations \cite{Basko_08}:
(i) $C_{2}$ rotation exchanging sublattices and (ii) mirror-reflection
$R_{x}$ leaving sublattices invariant. For example, the bubbles $M_{00,0a}(0,\omega)$,
encoding charge-density--spin-density-type responses ($a=x,y,z$),
vanish identically because the associated vertices (``$00$'' and
``$0a$'') transform differently under at least one unitary symmetry.
Overall, there are 64 non-zero bubbles in the long-wavelength $\mathbf{q}\rightarrow0$
limit. The symmetry classification is summarized in Table \ref{tab:Parity-of-Dirac}.

We now turn to the spin--charge eigenmodes sustained by the system.
The first step is to compute the gradient expansion of Eq. (\ref{eq:diff_Ham_exp}).
Working in the diffusive regime ($\epsilon\tau\gg1\gg\lambda\tau$)
greatly simplifies matters due to many bubbles being parametrically
small. To leading order in $v|\mathbf{q}|$ and $\omega\tau$, we
find 
\begin{equation}
\mathcal{H}_{\mathcal{D}}(\mathbf{q},\omega)\simeq\imath\omega Q-\imath\mathbf{P}\cdot\mathbf{q}+L\,,\label{eq:diff_Ham_small_expansion}
\end{equation}
where $Q$, $P_{i}$ and $L$ are $16\times16$ matrices given by
$Q=-\imath(\nabla_{\omega}\mathcal{H}_{\mathcal{D}}(0,\omega))_{\mathbf{\omega}=0}$,
$\mathbf{P}=\imath(\partial_{\mathbf{q}}\mathcal{H}_{\mathcal{D}}(\mathbf{q},0))_{\mathbf{q}=0}$
and $L=\mathcal{H}_{\mathcal{D}}(0,0)$. The calculation of these
matrices is rather cumbersome, yielding unwieldy expressions for the
matrices $Q=Q(\varepsilon,\lambda,\tau)$, $P_{i}=P_{i}(\varepsilon,\lambda,\tau)$
and $L=L(\varepsilon,\lambda,\tau)$. We therefore provide the explicit
form of the Eq. (\ref{eq:diff_Ham_small_expansion}) in the Appendix,
from which $Q$, $P_{i}$ and $L$ can be inferred. 

\begin{figure}
\begin{centering}
\includegraphics[width=0.4\columnwidth]{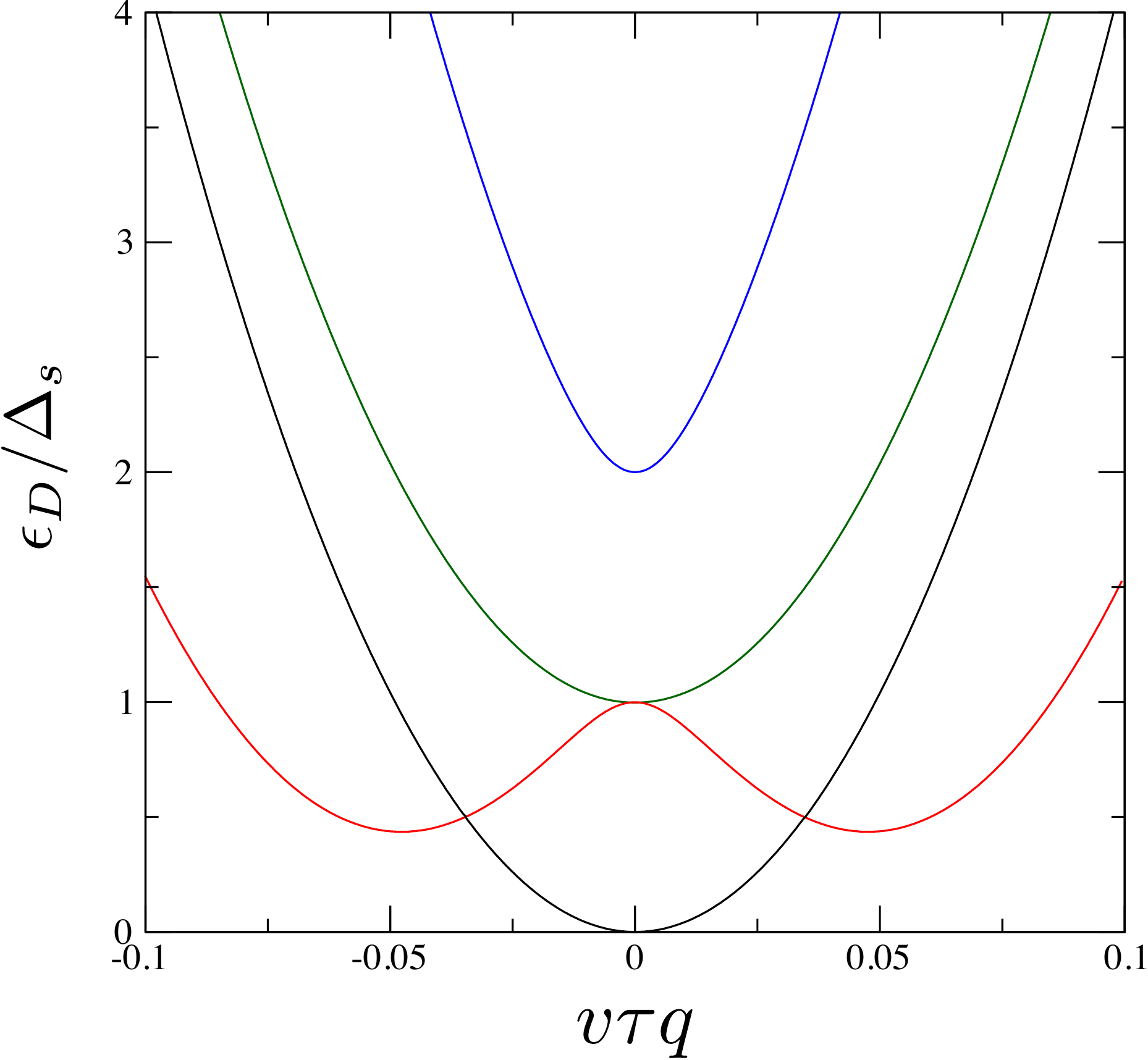}
\par\end{centering}
\caption{\label{fig:EigenModes}Spectrum of the \emph{diffuson} Hamiltonian
in units of spin gap $\Delta_{s}$. Only low-lying states are shown.
Parameters: $\varepsilon=0.3$ eV, $\lambda=0.1$ meV and $\eta=1/2\tau=2$
meV. The curves are obtained numerically from $\epsilon_{D}(q)=\textrm{Re}\,\lambda(q,0)$
with $q=|\mathbf{q}|$, where $\lambda(q,\omega)$ are the complex
eigenvalues of the \emph{diffuson} Hamiltonian matrix {[}Eq.\,(\ref{eq:diff_Ham_exp}){]}.}
\end{figure}

The \emph{diffuson} Hamiltonian in Eq. (\ref{eq:diff_Ham_small_expansion})
is linear in the wavevector $\mathbf{q}$ due to the Dirac nature
of the low-lying excitations in graphene heterostructures. The \emph{diffuson}
spectrum at low energies is shown in Fig. \ref{fig:EigenModes}. The
quadratic dispersion of the gapless mode follows from the diffusive
pole structure of the density--density response, i.e. $[\mathcal{R}_{00,00}(\mathbf{q},0)]^{-1}\sim(v\tau|\mathbf{q}|)^{2}$.
For $\mathbf{q}\neq0$, this mode is an admixture of charge $N$,
spin current $\mathcal{J}_{i}^{a}$ and spin polarization $S^{x,y}$
fluctuations. The gapped states, on the other hand, describe eigenmodes
of the nonequilibrium spin polarization $S^{a}$ \cite{RevModPhys.89.011001}.
Note that two of these states ($S^{x,y}$) are degenerate at $\mathbf{q}=0$
due to the rotational ($C_{v\infty}$) symmetry of the single-particle
Hamiltonian {[}Eq. (\ref{eq:Hamiltonian}){]}. Their gaps at $\mathbf{q}=0$
are given by $\Delta_{s}\equiv\Delta_{\parallel}\simeq(\lambda/\eta)^{2}$
for the $S^{x,y}$- and $\Delta_{\perp}\simeq2\Delta_{\parallel}$
for the $S^{z}$-mode. The twice as fast dephasing of out-of-plane
spin fluctuations is a fingerprint of BR SOC, a feature which is known
to survive even in the strong (unitary) scattering regime \cite{BR_Graphene_Offidani_MDPI_18}. 

The low-lying modes shown in Fig. \ref{fig:EigenModes} are remarkably
similar to those of a BR-coupled 2D electron gas \cite{Wenk_10}.
We verified that the \emph{diffuson} spectra of the two Rashba models
can be exactly mapped onto each other, despite their distinctly different
\emph{diffuson} Hamiltonians (i.e. $4\times4$ Schrödinger-like for
2D electron gases and 1$6\times16$ Dirac-like for graphene). Defining
$\xi=2v|\mathbf{q}|\tau$, the eigenvalues of $\mathcal{\mathcal{H}_{\mathcal{D}}}(\mathbf{q},0)$
in the limit $v|\mathbf{q}|\ll1$ are given by $\epsilon_{D}^{0}=\xi^{2}$,
$\epsilon_{D}^{1}=\xi^{2}+\Delta_{s}$ and $\epsilon_{D}^{\pm}=\xi^{2}+\Delta_{s}\left(\frac{3}{2}\pm\frac{1}{2}\sqrt{1+16\xi^{2}/\Delta_{s}}\right)$.
The Rashba \emph{diffuson} eigenvalues derived by Wenk\emph{ }et.
al. \cite{Wenk_10} are recovered by letting $\lambda/v\rightarrow2m_{e}\alpha_{2}$,
where $m_{e}$ is the effective electron mass and $\alpha_{2}$ the
Rashba parameter. The existence of such a mapping reflects the same
basic spin-relaxation (Dyakonov-Perel) mechanism at work. Indeed,
our findings put on a firm ground previous heuristic arguments \cite{BR_Graphene_Milletari_17,BR_Graphene_Offidani_MDPI_18}
for the equivalence of two models  in the weak SOC regime. The remaining
$16-4=12$ modes in the Dirac--Rashba model are characterized by
very large gaps ($\Delta_{s}^{+}\gtrsim1/2\gg\Delta_{s}$), and as
such play no role in the diffusive regime. 

A comment is in order regarding the validity of our assumptions in
the light of recent findings. In the example of Fig. \ref{fig:EigenModes},
we considered a small BR coupling of only $0.1$ meV, which is in
line with density functional theory calculations for clean graphene/group
VI dichalcogenide heterostructures \cite{Wang_14,Gmitra_16}. On the
other hand, the semi-empirical Slater-Koster parametrization of Ref.
\cite{Cysne_18}, as well as early magnetotransport measurements \cite{Yang_2016,Yang_17,Volkl_17,Zihlmann_18,Wakamura_18},
have found much higher proximity-induced SOC (up to $\approx10$ meV).
Furthermore, a recent joint theory-experiment study suggests that
the interfacial Rashba coupling can be made as large as 100 meV, by
placing a graphene flake on top of a metallic substrate with a suitable
work function mismatch \cite{Li_2020}. The theory developed in this
work is expected to remain accurate provided that the Dirac bands
remain intact (so that the low-energy picture in Fig. \ref{fig:schematic}
is justified) and the system is sufficiently disordered so that $\lambda\tau\ll1$. 

\subsection{Unified coupled spin-charge drift-diffusion equations\label{subsec:2_unified-coarse_grained_dynamics} }

The quantum kinetic equation governing the one-particle reduced density
matrix in the large distance and long time limits is obtained after
an inverse Fourier transform ($-\imath\omega\rightarrow\partial_{t}$
and $\imath q_{i}\rightarrow\nabla_{i}$) of Eq. (\ref{eq:diff_Ham_small_expansion})
as
\begin{equation}
\mathcal{H}_{\mathcal{D}}\cdot\delta\vec{\rho}(\mathbf{x},t)=\left(-Q\frac{\partial}{\partial t}-\sum_{i=x,y}P_{i}\nabla_{i}+L\right)\cdot\delta\vec{\rho}(\mathbf{x},t)=v\nu_{0}\partial_{t}\vec{\mathcal{A}}_{\textrm{ext}}(\mathbf{x},t)\,.\label{eq:grad_expansion}
\end{equation}
Transport equations for the coarse-grained variables $N(\mathbf{x},t)=-\delta\rho_{00}(\mathbf{x},t)$,
$S^{a}(\mathbf{x},t)=\frac{1}{2}\delta\rho_{0a}(\mathbf{x},t)$, $J_{i}(\mathbf{x},t)=-v\delta\rho_{i0}(\mathbf{x},t)$
and $\mathcal{J}_{i}^{a}(\mathbf{x},t)=\frac{v}{2}\delta\rho_{ia}(\mathbf{x},t)$
can now be derived by replacing the $16\times16$ matrices $Q=Q(\varepsilon,\lambda,\tau)$,
$P_{i}=P_{i}(\varepsilon,\lambda,\tau)$ and $L=L(\varepsilon,\lambda,\tau)$
with their explicit forms (see Appendix). To leading order in $1/\varepsilon\tau$
and $\lambda\tau$, we obtain
\begin{equation}
\partial_{t}N+\nabla\cdot J=0\,,\label{eq:SDE1}
\end{equation}
\begin{equation}
\partial_{t}S^{a}+\nabla\cdot\mathcal{J}^{a}=\frac{2\lambda}{v}\,\Lambda_{c}^{ab}\mathcal{J}_{b}^{c}+\frac{\nu_{0}}{2}\mathcal{B}^{a}\,,\label{eq:SDE2}
\end{equation}
\begin{equation}
\partial_{t}J_{i}+\frac{v^{2}}{2}\nabla_{i}N=-\frac{J_{i}}{2\tau}-\frac{4\lambda^{2}}{\epsilon}\epsilon_{ij}\mathcal{J}_{j}^{z}+\frac{\nu_{0}}{2}v^{2}E_{i}\,,\label{eq:SDE3}
\end{equation}
\begin{equation}
\partial_{t}\mathcal{J}_{i}^{a}+\frac{v^{2}}{2}\nabla_{i}S^{a}+v\lambda\tau\,\Gamma_{ic}^{ab}\nabla_{b}\mathcal{J}_{i}^{c}=-\frac{\mathcal{J}_{i}^{a}}{2\tau}+v\lambda\varOmega_{ic}^{a}S^{c}+\lambda^{2}\tau\,\varUpsilon_{ic}^{ab}\mathcal{J}_{b}^{c}-\delta_{az}\epsilon_{ij}\frac{\lambda^{2}}{\varepsilon}J_{j}\,,\label{eq:SDE4}
\end{equation}
where $E_{i}=-\partial_{t}\mathcal{A}_{i0}^{\textrm{ext}}$ is the
externally applied electric field and $\mathcal{B}^{a}=(1/v)\thinspace\partial_{t}\mathcal{A}_{0a}^{\textrm{ext}}(\mathbf{x},t)$
is the external Zeeman field (``spin injection field'') that induces
a nonequilibrium spin density \cite{SOC_2DEG_Ka_Raimondi_Vignale_14b}.
The coefficients $\Lambda_{c}^{ab},\Gamma_{ic}^{ab},\varOmega_{ic}^{a}$
and $\varUpsilon_{ic}^{ab}$ are listed in Table \ref{tab:Coefficients}
and $\epsilon_{ij}$ denotes the rank-2 Levi-Civita symbol ($i,j=x,y$).
Different perturbations (e.g. a spin-dependent electric field $\mathcal{E}_{i}^{a}=-\partial_{t}\mathcal{A}_{ia}$)
can be easily incorporated via a suitable parameterization of the
generalized external vector potential entering Eq. (\ref{eq:grad_expansion}).

The drift--diffusion equations (\ref{eq:SDE1})-(\ref{eq:SDE4})
are the main results of this work. Equations (\ref{eq:SDE1}) and
(\ref{eq:SDE2}) are generalized continuity relations and Eqs. (\ref{eq:SDE3})
and (\ref{eq:SDE4}) express the time evolution of the currents as
a sum of drift, diffusion, spin precession and spin--charge conversion
processes. Standard constitutive relations

\begin{table}[b]
\begin{centering}
\begin{tabular}{|c|c|c|}
\hline 
$\Lambda$ & spin density precession & $\Lambda_{x}^{zx}=\Lambda_{y}^{zy}=-1\,,$$\;$$\Lambda_{z}^{xx}=\Lambda_{z}^{yy}=1$\tabularnewline
\hline 
$\Gamma$ & spin current transfer & $\Gamma_{xz}^{xx}=\Gamma_{yz}^{yy}=1,$$\;$$\Gamma_{xx}^{zx}=\Gamma_{xy}^{zy}=\Gamma_{yx}^{zx}=\Gamma_{yy}^{zy}=-1,$
$\Gamma_{xz}^{yy}=\Gamma_{yz}^{xx}=1$\tabularnewline
\hline 
$\varOmega$ & spin current precession & $\varOmega_{xz}^{x}=\varOmega_{yz}^{y}=1,$$\;$$\varOmega_{xx}^{z}=\varOmega_{yy}^{z}=-1$\tabularnewline
\hline 
$\varUpsilon$ & spin current swapping & $\varUpsilon_{xy}^{xy}=\varUpsilon_{yx}^{yx}=1$, $\;$$\varUpsilon_{xx}^{yy}=\varUpsilon_{yy}^{xx}=-1$\tabularnewline
\hline 
\end{tabular}
\par\end{centering}
\caption{\label{tab:Coefficients}Coefficients in the coupled spin--charge
drift--diffusion equations. Only nonzero components are listed. }
\end{table}

\begin{equation}
J_{i}=-D\,\nabla_{i}N-\frac{8\tau\lambda^{2}}{\varepsilon}\epsilon_{ij}\mathcal{J}_{j}^{z}+\nu_{0}\tau v^{2}E_{i}\,,\label{eq:DDE3}
\end{equation}
\begin{equation}
\mathcal{J}_{i}^{a}=-D\nabla_{i}S^{a}-2v\lambda\tau^{2}\,\Gamma_{ic}^{ab}\nabla_{b}\mathcal{J}_{i}^{c}+2v\lambda\tau\varOmega_{ic}^{a}S^{c}+2\lambda^{2}\tau^{2}\,\varUpsilon_{ic}^{ab}\mathcal{J}_{b}^{c}-\delta_{a,z}\epsilon_{ij}\frac{2\tau\lambda^{2}}{\varepsilon}J_{j}\,,\label{eq:DDE4}
\end{equation}
where $D=v^{2}\tau$ is the diffusion constant \cite{note:3}, hold
to good accuracy insofar as $\omega\ll\lambda$. Sublattice-staggered
charge $N_{s}$ and spin densities $S_{s}^{a}$ (see Table \ref{tab:Parity-of-Dirac})
are conspicuously absent from these relations. In terms of the underlying
\emph{diffuson} Hamiltonian, these observables are linked to dispersionless
modes with large gaps and are thus effectively decoupled from the
low-energy dynamics. We expect such terms to play a role in graphene
heterostructures with broken sublattice symmetry \cite{Cummings2017,Manuel2018,BR_Graphene_Milletari_17},
which is beyond the scope of the this article. 

\subsection{Spin Hall and spin-galvanic effects: the DC regime \label{subsec:Spin-charge-conversion:-Onsager}}

Equations (\ref{eq:SDE1})--(\ref{eq:SDE4}) provide a physically
transparent scheme to interpret and predict a variety of SOC phenomena
of fundamental and technological relevance. Before discussing new
applications of the formalism, we briefly revisit two well established
results for BR-coupled graphene \cite{BR_Graphene_Offidani_17,BR_Graphene_Milletari_17}.
We start with the SHE \cite{Hirsch_99}, i.e. the appearance of a
transverse spin current upon application of a DC charge current, first
observed in semiconductors \cite{Kato1910,Wunderlich_05}. In graphene
with random SOC (e.g. induced by dilute adatoms), a robust SHE can
be induced via resonant skew scattering \cite{Ferreira_14,Balakrishnan_14,Pachoud_14,Yang_16,Huang_16}.
In contrast, for graphene systems with a spatially uniform BR effect,
the SHE is strictly vanishing unless supplemented with proximity-induced
spin--valley coupling or spin-dependent disorder.

Detailed information on the SHE can be obtained from the drift--diffusion
equations with little effort. The so-called \emph{intrinsic} spin
Hall angle $\theta_{\textrm{sH}}^{\textrm{int}}=2\tau\lambda^{2}/\varepsilon$,
which appears explicitly in Eqs. (\ref{eq:DDE3})--(\ref{eq:DDE4}),
diverges in the clean limit ($\tau\rightarrow\infty$) and does not
correspond \emph{per se} to a steady-state transport quantity. The
actual spin Hall angle, defined as the ratio between near-equilibrium
spin Hall current and applied charge current,
\begin{equation}
\theta_{\textrm{sH}}(\omega):=2\epsilon_{ij}\left.\frac{\mathcal{J}_{i}^{z}}{J_{j}}\right|_{\mathbf{q}=0,\mathcal{B}^{a}=0}=\;\theta_{\textrm{sH}}^{\textrm{int}}+\theta_{\textrm{sH}}^{\textrm{dis}}(\omega)\,,\label{eq:total_sH_angle}
\end{equation}
is obtained by solving the system of coupled equations (\ref{eq:SDE1})--(\ref{eq:SDE4})
and receives important disorder corrections even in the clean limit.
The ``SHE cancellation'' in the DC limit $\theta_{\textrm{sH}}^{\textrm{dis}}(0)=-\theta_{\textrm{sH}}^{\textrm{int}}$
is a fundamental consequence of SU(2)-spin covariance of pure Rashba
models as shown by Dimitrova \cite{Dimitrova_05}. This result can
also be interpreted as the unavoidable outcome for a 2D system with
an isotropic and fully in-plane spin texture. Because the electronic
states are admixtures of orthogonal spin states, phase shifts experienced
by the spin-up and spin-down components of scattered wavefunctions
from scalar impurities cannot be distinguished, implying the absence
of skew scattering \cite{BR_Graphene_Milletari_17}. As discussed
in Sec. \ref{Sub_sec_applications}, a robust SHE nevertheless takes
place at finite frequencies (i.e. an optical SHE) or when the system
is perturbed by a spin-injection field. 

Also of interest is the inverse spin-galvanic effect (ISGE), whereby
an applied current magnetizes the conduction electrons, thus generating
a net spin polarization density \cite{Aranov_89,BR_Edelstein90}.
Its microscopic origin lies in the spin--momentum locking of Bloch
eigenstates caused by the BR effect {[}Fig. \ref{fig:schematic} (c){]}.
While the equilibrium spin polarization averaged over the Fermi surface
in a nonmagnetic system must vanish identically {[}Eq. (\ref{eq:SP}){]},
an external electric field effectively breaks the time-reversal symmetry,
by causing an imbalance  the occupation of states with opposite momenta,
 which allows the build up of a net transverse spin polarization.
The ISGE efficiency can be easily read out from Eq. (\ref{eq:DDE4})
by replacing the pure spin current by its steady-state value in the
minimal model, i.e. $\mathcal{J}_{i}^{a}=0$. The charge-to-spin conversion
efficiency is obtained as:
\begin{equation}
\kappa_{ij}:=2v\left.\frac{S^{i}}{J_{j}}\right|_{\mathbf{q}=0,\mathcal{B}^{a}=0}=\epsilon_{ij}\frac{1}{\lambda\tau}\theta_{\textrm{sH}}^{\textrm{int}}=\epsilon_{ij}\frac{2\lambda}{\varepsilon}\,.\label{eq:REE}
\end{equation}
This relation (first derived in Ref. \cite{BR_Graphene_Offidani_17})
discloses an optimal spin-charge conversion efficiency at the spin-gap
edge, i.e. $\kappa_{xy}(\varepsilon=2\lambda)=1$. The ISGE efficiency
parameter decays algebraically with the energy of charge carriers,
which makes the effect detectable at room temperature over a wide
range of charge carrier densities. The robust ISGE in graphene with
BR effect has been observed in a recent series of experiments \cite{Ghiasi_19,Li_2020,Benitez_20,Hoque_21}.

\subsection{Application: Optical spin Hall and spin galvanic effects \label{Sub_sec_applications}}

As a novel application of the formalism, we derive the optical response
of BR-coupled graphene. To this end, we solve the charge--spin drift--diffusion
equations {[}Eqs.\,(\ref{eq:SDE2})-(\ref{eq:SDE4}){]} subject to
time-dependent electric and spin-injection fields, with Fourier transforms
$E_{i}(\omega)$ ($i=x,y$) and $\mathcal{B}_{a}(\omega)$ ($a=x,y,z$),
respectively. In the Dyakonov-Perel regime with $\omega\tau\ll\lambda\tau\ll1\ll\epsilon\tau$,
the macroscopic observables of interest are found, after tedious but
straightforward calculations, to be 

\begin{align}
J_{i}(\omega) & =\frac{g_{v}\varepsilon\tau}{\pi}\,\left[\frac{E_{i}(\omega)}{1-2\imath\omega\tau}+\epsilon_{ij}\frac{2\lambda}{v\varepsilon}\frac{\mathcal{B}_{j}(\omega)}{\left(1-\imath\omega\tau_{\parallel}\right)\left(1-2\imath\omega\tau\right)}\right]\,,\label{eq:AC_charge_current}\\
\nonumber \\
S_{a}(\omega) & =\frac{g_{v}\varepsilon\tau_{s,a}}{2\pi v^{2}}\,\left[\epsilon_{jaz}\,\frac{8v\lambda^{3}\tau^{2}}{\varepsilon}\frac{E_{j}(\omega)}{\left(1-2\imath\omega\tau\right)\left(1-\imath\omega\tau_{s,a}\right)}+\frac{\mathcal{B}_{a}(\omega)}{1-\imath\omega\tau_{s,a}}\right]\,,\label{eq:AC_spin_density}\\
\nonumber \\
\mathcal{J}_{i}^{z}(\omega) & =\frac{g_{v}\varepsilon\tau_{ss}^{z}}{2\pi v}\,\left[\imath v\epsilon_{ij}\frac{\omega\tau}{\varepsilon\tau_{ss}^{z}}\frac{E_{j}(\omega)}{\left(1-2\imath\omega\tau\right)\left(1-\imath\omega\tau_{\parallel}\right)}-\frac{\mathcal{B}_{i}(\omega)}{1-\imath\omega\tau_{\parallel}}\right]\,,\label{eq:AC_spin_current}\\
\nonumber \\
\mathcal{J}_{i}^{i}(\omega) & =\frac{g_{v}\varepsilon\tau_{ss}^{i}}{2\pi v}\,\frac{\mathcal{B}_{z}(\omega)}{1-\imath\omega\tau_{\perp}}\,,\label{eq:AC_spin_current_ii}
\end{align}
where the factor of $g_{v}=2$ accounts for valley degeneracy. Furthermore,
$\epsilon_{zij}$ denotes the rank-3 Levi-Civita symbol, $\tau_{ss}^{i=x,y}=\tau_{ss}^{z}/2$
with $\tau_{ss}^{z}=1/2\lambda$, and $\tau_{s,a}$ is the Dyakonov-Perel
relaxation time introduced earlier {[}here, $\tau_{s,(x,y)}\equiv\tau_{\parallel}$
and $\tau_{s,z}\equiv\tau_{\perp}=\tau_{\parallel}/2$, with $\tau_{\parallel}=(4\lambda^{2}\tau)^{-1}${]}.

Equations (\ref{eq:AC_charge_current})--(\ref{eq:AC_spin_current_ii})
reveal several interesting features of the Dirac--Rashba model. First,
a finite  SHE is established in the presence of a time-dependent electric
field {[}Eq. (\ref{eq:AC_spin_current}){]}. Second, a spin-injection
field gives rise to a charge current (i.e. direct spin galvanic effect)
{[}Eq. (\ref{eq:AC_charge_current}){]} and a nonequilibrium spin
polarization {[}Eq. (\ref{eq:AC_spin_density}){]}. Moreover, a pure
spin current is induced by an applied spin-injection field {[}Eqs.
(\ref{eq:AC_spin_current})--(\ref{eq:AC_spin_current_ii}){]}. The
induced spin-current density is polarized transversely to the spin-injection
field. With the exception of the SHE, all such effects are present
in the DC limit.

The linear response functions to external fields can be readily obtained
from Eqs. (\ref{eq:AC_charge_current})--(\ref{eq:AC_spin_current_ii}).
For example, the optical conductivity {[}$\sigma_{ij}(\omega)=\delta J_{i}(\omega)/\delta E_{j}(\omega)${]},
spin-galvanic susceptibility {[}$\chi_{ij}(\omega)=\delta S_{i}(\omega)/\delta E_{j}(\omega)${]}
and spin Hall conductivity {[}$\sigma_{yx}^{z}(\omega)=\delta\mathcal{J}_{y}^{z}(\omega)/\delta E_{x}(\omega)${]}
read as
\begin{equation}
\sigma_{ii}(\omega)=2g_{v}\,\frac{\varepsilon\tau}{1-2\imath\omega\tau}\,,\qquad(\mathrm{units\,of\,}e^{2}/h)\label{eq:AC_conductivity}
\end{equation}
\begin{equation}
\chi_{ij}(\omega)\simeq-\frac{2g_{v}\epsilon_{ij}}{v}\,\frac{\lambda\tau}{1-\imath\omega\tau_{\parallel}}\,,\qquad(\mathrm{units\,of\,}e/2\pi)\label{eq:AC_Edelstein_susceptibility}
\end{equation}
\begin{equation}
\sigma_{ij}^{z}(\omega)\simeq-\imath\epsilon_{ij}g_{v}\,\frac{\omega\tau}{1-\imath\omega\tau_{\parallel}}\,,\qquad(\mathrm{units\,of\,}e/2\pi)\label{eq:AC_SH_conductivity}
\end{equation}
 where subleading corrections of order $\omega\tau$ have been neglected
for simplicity. We note that the expression for $\sigma_{xx}(\omega)$
coincides with the familiar Drude model result, which is valid in
the semiclassical regime with $\varepsilon\tau\gg1$ \cite{Bludov_13}.
More interestingly, the optical spin-galvanic susceptibility {[}Eq.\,(\ref{eq:AC_Edelstein_susceptibility}){]}
generalizes the findings of Ref. \cite{BR_Graphene_Offidani_17} to
an external electric field with nonzero frequency. Here, $\omega\tau_{\parallel}$
emerges as an important parameter that governs the imaginary part
of the response function, whereas the physics of the DC regime reflects
the average spin-precession angle experienced between consecutive
scattering events (i.e. $\theta_{p}=\lambda\tau$). 

\section{Summary and outlook \label{sec:Summary-and-outlook}}

We derived a quantum kinetic equation and the associated set of coupled
spin--charge linear transport equations that govern the dynamics
of Rashba-coupled 2D Dirac fermions in graphene proximitized by high-SOC
materials. These equations, which are valid in the presence of arbitrary
external fields, provide a quantitative description of rich interlinked
spin--orbit scattering phenomena characteristic of 2D systems with
broken inversion symmetry, including Dyakonov-Perel-type spin relaxation,
direct and inverse DC spin-galvanic and optical spin Hall effects. 

The distinctive feature of the SO(5) algebraic approach formulated
in this work is that the exact large distance and long time behavior
of the linear response one-particle density matrix, and thus also
the expectation value of any local observable, is uniquely determined
by a generalized \emph{inverse} \emph{diffuson} matrix (i.e. a \emph{diffuson}
Hamiltonian) that spans the full vector space of a 16-dimensional
Clifford algebra. This is to be contrasted with the familiar SU(2)
approach for two-dimensional electron gases \cite{2DEG_SOC_Burkov_04,SOC_2DEG_Ka_Raimondi_Vignale_14a},
whose Fourier-space \emph{diffuson} operators are $4\times4$ matrices
restricted to the space of charge and spin-polarization densities.
The enlarged ($16\times16$) \emph{diffuson} Hamiltonian derived here
emphasizes the manifestation of entanglement between the spin and
pseudospin (sublattice) degrees of freedom that is ubiquitous across
van der Waals materials. Furthermore, it provides direct access to
the time evolution of all thermodynamic macroscopic observables, including
spin-current density, spin-polarization density and sublattice-staggered
densities. 

This work opens up a number of avenues that can be explored in the
framework introduced here. These range from the exploration of nonequilibrium
opto-spintronic phenomena in group-VI dichalcogenide monolayers and
van der Waals heterostructures with sizable spin--valley coupling,
to the role played by spin--orbit-active impurities in  spin dynamics
and spin--charge conversion effects. In particular, asymmetric scattering
precession \cite{Huang_16} and skew scattering effects can be systematically
explored by means of a nonperturbative $T$-matrix ladder scheme that
resums all single-impurity scattering diagrams \cite{BR_Graphene_Offidani_17,BR_Graphene_Offidani_MDPI_18,Sousa_2020}.
Such an extension of our diagrammatic treatment would be helpful in
understanding the emergent transport physics of 2D van der Waals materials
with broken sublattice symmetry, where the presence of non-coplanar
$\mathbf{k}$-space spin textures at low energies is known to enable
robust spin Hall effects irrespective of the type of impurities \cite{BR_Graphene_Milletari_17,Offidani_Ferreira_18},
in addition to strongly modifying the spin dynamics \cite{Cummings2017,Manuel2018}.
Moreover, our formalism could be employed to investigate how proximity-induced
SOC affects the nonlocal resistance in Hanle-type spin precession
experiments beyond its impact on the spin lifetimes. This could be
examined by deriving a generalized spin diffusion equation accounting
for the interplay of spin-valley coupling, intervalley scattering
and the Bychkov-Rashba effect.

\subsection*{Data statement}

This publication is theoretical work that does not require supporting
research data.

\subsection*{Acknowledgements}

The author thanks D. Perkins for proofreading the final version of
the manuscript. This work was funded by the Royal Society through
a Royal Society University Research Fellowship (Grant No. URF\textbackslash R\textbackslash 191021). 

\section{Appendix}

The \emph{diffuson} Hamiltonian in the standard weak SOC regime with
$\epsilon\tau\gg1$ reads as 
\begin{equation}
\mathcal{H}_{\mathcal{D}}(\mathbf{q},\omega)=\imath\omega Q-\imath P\cdot\mathbf{q}+L\simeq\frac{1}{\tau}\left(\begin{array}{cc}
\left(\begin{array}{cc}
\mathcal{A}_{11} & \mathcal{A}_{12}\\
\mathcal{A}_{12} & \mathcal{A}_{22}
\end{array}\right) & \left(\begin{array}{cc}
\mathcal{C}_{11} & \mathbb{O}\\
\mathcal{C}_{21} & \mathbb{O}
\end{array}\right)\\
\left(\begin{array}{cc}
\mathcal{C}_{11} & -\mathcal{C}_{21}\\
\mathbb{O} & \mathbb{O}
\end{array}\right) & \left(\begin{array}{cc}
\mathcal{B}_{11} & \mathbb{O}\\
\mathbb{O} & \mathcal{B}_{22}
\end{array}\right)
\end{array}\right)\,,\label{eq:H_Q_P_L}
\end{equation}
where $\mathbb{O}$ is the null matrix, $\mathcal{A}_{11}=-\imath\omega\tau\,\mathbb{I}$,
\begin{align}
\mathcal{A}_{12}=\left(\begin{array}{cccc}
\imath vq_{x}\tau & 0 & 0 & 0\\
0 & \imath vq_{x}\tau & 0 & -2\lambda\tau\\
0 & 0 & \imath vq_{x}\tau & 0\\
0 & 2\lambda\tau & 0 & \imath vq_{x}\tau
\end{array}\right)\,,\label{eq:A12}\\
\nonumber \\
\mathcal{A}_{22}=\left(\begin{array}{cccc}
1-2\imath\omega\tau & 0 & 0 & 0\\
0 & 1-2\imath\omega\tau & 0 & 2\imath vq_{x}\lambda\tau^{2}\\
0 & 0 & 1-2\imath\omega\tau & 2\imath vq_{y}\lambda\tau^{2}\\
0 & -2\imath vq_{x}\lambda\tau^{2} & -2\imath vq_{y}\lambda\tau^{2} & 1-2\imath\omega\tau
\end{array}\right)\,,\label{eq:A22}\\
\nonumber 
\end{align}
\begin{align}
\mathcal{B}_{11}=\left(\begin{array}{cccc}
1-2\imath\omega\tau & 0 & 0 & 0\\
0 & 1-2\imath\omega\tau & 0 & 2\imath vq_{x}\lambda\tau^{2}\\
0 & 0 & 1-2\imath\omega\tau & 2\imath vq_{y}\lambda\tau^{2}\\
0 & -2\imath vq_{x}\lambda\tau^{2} & -2\imath vq_{y}\lambda\tau^{2} & 1-2\imath\omega\tau
\end{array}\right)\,,\label{eq:B11}
\end{align}
\begin{align}
\mathcal{C}_{11}=\left(\begin{array}{cccc}
\imath vq_{y}\tau & 0 & 0 & 0\\
0 & \imath vq_{y}\tau & 0 & 0\\
0 & 0 & \imath vq_{y}\tau & -2\lambda\tau\\
0 & 0 & 2\lambda\tau & \imath vq_{y}\tau
\end{array}\right)\,,\label{eq:C11}\\
\nonumber \\
\mathcal{C}_{21}=\left(\begin{array}{cccc}
0 & 0 & 0 & -\frac{4\lambda^{2}\tau}{\varepsilon}\\
0 & 0 & -2\tau^{2}\lambda^{2} & 0\\
0 & 2\tau^{2}\lambda^{2} & 0 & 0\\
-\frac{4\lambda^{2}\tau}{\varepsilon} & 0 & 0 & 0
\end{array}\right)\,,\label{eq:C12}
\end{align}
and $\mathcal{B}_{22}=[-\pi\varepsilon\tau/\ln\left(v\Lambda/\varepsilon\right)+\imath\pi^{2}\tau\omega/(4\ln^{2}\left(v\Lambda/\varepsilon\right))]\,\mathbb{I}$.
Here, $\mathbb{I}$ denotes the $4\times4$ identity matrix and $\Lambda$
is a momentum cutoff used to regularize the integrals in Eq. (\ref{eq:bubbles}).

\medskip{}

\pagebreak{}

\bibliographystyle{unsrt}
\bibliography{paper}

\begin{thebibliography}{10}

\bibitem{RevModPhys.91.035004}
A.~Manchon, J.~\ifmmode~\check{Z}\else \v{Z}\fi{}elezn\'y, I.~M. Miron,
  T.~Jungwirth, J.~Sinova, A.~Thiaville, K.~Garello, and P.~Gambardella.
\newblock Current-induced spin-orbit torques in ferromagnetic and
  antiferromagnetic systems.
\newblock {\em Rev. Mod. Phys.}, 91:035004, Sep 2019.

\bibitem{RevModPhys.89.025006}
Frances Hellman, Axel Hoffmann, Yaroslav Tserkovnyak, Geoffrey S.~D. Beach,
  Eric~E. Fullerton, Chris Leighton, Allan~H. MacDonald, Daniel~C. Ralph,
  Dario~A. Arena, Hermann~A. D\"urr, Peter Fischer, Julie Grollier, Joseph~P.
  Heremans, Tomas Jungwirth, Alexey~V. Kimel, Bert Koopmans, Ilya~N.
  Krivorotov, Steven~J. May, Amanda~K. Petford-Long, James~M. Rondinelli, Nitin
  Samarth, Ivan~K. Schuller, Andrei~N. Slavin, Mark~D. Stiles, Oleg
  Tchernyshyov, Andr\'e Thiaville, and Barry~L. Zink.
\newblock Interface-induced phenomena in magnetism.
\newblock {\em Rev. Mod. Phys.}, 89:025006, Jun 2017.

\bibitem{RevModPhys.92.021003}
A.~Avsar, H.~Ochoa, F.~Guinea, B.~\"Ozyilmaz, B.~J. van Wees, and I.~J.
  Vera-Marun.
\newblock Colloquium: Spintronics in graphene and other two-dimensional
  materials.
\newblock {\em Rev. Mod. Phys.}, 92:021003, Jun 2020.

\bibitem{Sierra_21}
Juan~F. Sierra, Jaroslav Fabian, Roland~K. Kawakami, Stephan Roche, and
  Sergio~O. Valenzuela.
\newblock Van der waals heterostructures for spintronics and opto-spintronics.
\newblock {\em Nature Nanotechnology}, 16(8):856--868, 2021.

\bibitem{Tombros_2007}
Nikolaos Tombros, Csaba Jozsa, Mihaita Popinciuc, Harry~T. Jonkman, and Bart~J.
  van Wees.
\newblock Electronic spin transport and spin precession in single graphene
  layers at room temperature.
\newblock {\em Nature}, 448(7153):571--574, 2007.

\bibitem{Han_11}
Wei Han and R.~K. Kawakami.
\newblock Spin relaxation in single-layer and bilayer graphene.
\newblock {\em Phys. Rev. Lett.}, 107:047207, Jul 2011.

\bibitem{Zomer_12}
P.~J. Zomer, M.~H.~D. Guimar\~aes, N.~Tombros, and B.~J. van Wees.
\newblock Long-distance spin transport in high-mobility graphene on hexagonal
  boron nitride.
\newblock {\em Phys. Rev. B}, 86:161416, Oct 2012.

\bibitem{Kamalakar_15}
M.~Venkata Kamalakar, Christiaan Groenveld, Andr{\'e} Dankert, and Saroj~P.
  Dash.
\newblock Long distance spin communication in chemical vapour deposited
  graphene.
\newblock {\em Nature Communications}, 6(1):6766, 2015.

\bibitem{Yan_16}
W.~Yan, L.~C. Phillips, M.~Barbone, S.~J. H\"am\"al\"ainen, A.~Lombardo,
  M.~Ghidini, X.~Moya, F.~Maccherozzi, S.~van Dijken, S.~S. Dhesi, A.~C.
  Ferrari, and N.~D. Mathur.
\newblock Long spin diffusion length in few-layer graphene flakes.
\newblock {\em Phys. Rev. Lett.}, 117:147201, Sep 2016.

\bibitem{Gebeyehu_2019}
Z~M Gebeyehu, S~Parui, J~F Sierra, M~Timmermans, M~J Esplandiu, S~Brems,
  C~Huyghebaert, K~Garello, M~V Costache, and S~O Valenzuela.
\newblock Spin communication over 30 $\mathrm{\mu}$ m long channels of chemical
  vapor deposited graphene on {SiO} 2.
\newblock {\em 2D Materials}, 6(3):034003, may 2019.

\bibitem{Sepioni_10}
M.~Sepioni, R.~R. Nair, S.~Rablen, J.~Narayanan, F.~Tuna, R.~Winpenny, A.~K.
  Geim, and I.~V. Grigorieva.
\newblock Limits on intrinsic magnetism in graphene.
\newblock {\em Phys. Rev. Lett.}, 105:207205, Nov 2010.

\bibitem{SOC_Graphene_Sichau_19}
J.~Sichau, M.~Prada, T.~Anlauf, T.~J. Lyon, B.~Bosnjak, L.~Tiemann, and R.~H.
  Blick.
\newblock Resonance microwave measurements of an intrinsic spin-orbit coupling
  gap in graphene: A possible indication of a topological state.
\newblock {\em Phys. Rev. Lett.}, 122:046403, Feb 2019.

\bibitem{Ferreira_14}
Aires Ferreira, Tatiana~G. Rappoport, Miguel~A. Cazalilla, and A.~H.
  Castro~Neto.
\newblock Extrinsic spin hall effect induced by resonant skew scattering in
  graphene.
\newblock {\em Phys. Rev. Lett.}, 112:066601, Feb 2014.

\bibitem{Balakrishnan_14}
Jayakumar Balakrishnan, Gavin Kok~Wai Koon, Ahmet Avsar, Yuda Ho, Jong~Hak Lee,
  Manu Jaiswal, Seung-Jae Baeck, Jong-Hyun Ahn, Aires Ferreira, Miguel~A.
  Cazalilla, Antonio H.~Castro Neto, and Barbaros {\"O}zyilmaz.
\newblock Giant spin hall effect in graphene grown by chemical vapour
  deposition.
\newblock {\em Nature Communications}, 5(1):4748, 2014.

\bibitem{Avsar_14}
A.~Avsar, J.~Y. Tan, T.~Taychatanapat, J.~Balakrishnan, G.~K.~W. Koon, Y.~Yeo,
  J.~Lahiri, A.~Carvalho, A.~S. Rodin, E.~C.~T. O'Farrell, G.~Eda, A.~H.
  Castro~Neto, and B.~{\"O}zyilmaz.
\newblock Spin--orbit proximity effect in graphene.
\newblock {\em Nature Communications}, 5(1):4875, 2014.

\bibitem{Wang_14}
Zhe Wang, Dong-Keun Ki, Hua Chen, Helmuth Berger, Allan~H. MacDonald, and
  Alberto~F. Morpurgo.
\newblock Strong interface-induced spin--orbit interaction in graphene on ws2.
\newblock {\em Nature Communications}, 6(1):8339, 2015.

\bibitem{Wang_16}
Zhe Wang, Dong-Keun Ki, Jun~Yong Khoo, Diego Mauro, Helmuth Berger, Leonid~S.
  Levitov, and Alberto~F. Morpurgo.
\newblock Origin and magnitude of `designer' spin-orbit interaction in graphene
  on semiconducting transition metal dichalcogenides.
\newblock {\em Phys. Rev. X}, 6:041020, Oct 2016.

\bibitem{Gmitra_16}
Martin Gmitra, Denis Kochan, Petra H\"ogl, and Jaroslav Fabian.
\newblock Trivial and inverted dirac bands and the emergence of quantum spin
  hall states in graphene on transition-metal dichalcogenides.
\newblock {\em Phys. Rev. B}, 93:155104, Apr 2016.

\bibitem{Island_19}
J.~O. Island, X.~Cui, C.~Lewandowski, J.~Y. Khoo, E.~M. Spanton, H.~Zhou,
  D.~Rhodes, J.~C. Hone, T.~Taniguchi, K.~Watanabe, L.~S. Levitov, M.~P.
  Zaletel, and A.~F. Young.
\newblock Spin--orbit-driven band inversion in bilayer graphene by the van der
  waals proximity effect.
\newblock {\em Nature}, 571(7763):85--89, 2019.

\bibitem{Yang_2016}
Bowen Yang, Min-Feng Tu, Jeongwoo Kim, Yong Wu, Hui Wang, Jason Alicea, Ruqian
  Wu, Marc Bockrath, and Jing Shi.
\newblock Tunable spin{\textendash}orbit coupling and symmetry-protected edge
  states in graphene/{WS} 2.
\newblock {\em 2D Materials}, 3(3):031012, sep 2016.

\bibitem{Yang_17}
Bowen Yang, Mark Lohmann, David Barroso, Ingrid Liao, Zhisheng Lin, Yawen Liu,
  Ludwig Bartels, Kenji Watanabe, Takashi Taniguchi, and Jing Shi.
\newblock Strong electron-hole symmetric rashba spin-orbit coupling in
  graphene/monolayer transition metal dichalcogenide heterostructures.
\newblock {\em Phys. Rev. B}, 96:041409, Jul 2017.

\bibitem{Volkl_17}
Tobias V\"olkl, Tobias Rockinger, Martin Drienovsky, Kenji Watanabe, Takashi
  Taniguchi, Dieter Weiss, and Jonathan Eroms.
\newblock Magnetotransport in heterostructures of transition metal
  dichalcogenides and graphene.
\newblock {\em Phys. Rev. B}, 96:125405, Sep 2017.

\bibitem{Zihlmann_18}
Simon Zihlmann, Aron~W. Cummings, Jose~H. Garcia, M\'at\'e Kedves, Kenji
  Watanabe, Takashi Taniguchi, Christian Sch\"onenberger, and P\'eter Makk.
\newblock Large spin relaxation anisotropy and valley-zeeman spin-orbit
  coupling in ${\mathrm{wse}}_{2}$/graphene/$h$-bn heterostructures.
\newblock {\em Phys. Rev. B}, 97:075434, Feb 2018.

\bibitem{Wakamura_18}
T.~Wakamura, F.~Reale, P.~Palczynski, S.~Gu\'eron, C.~Mattevi, and H.~Bouchiat.
\newblock Strong anisotropic spin-orbit interaction induced in graphene by
  monolayer ${\mathrm{ws}}_{2}$.
\newblock {\em Phys. Rev. Lett.}, 120:106802, Mar 2018.

\bibitem{BR_Graphene_Rashba_09}
Emmanuel~I. Rashba.
\newblock Graphene with structure-induced spin-orbit coupling: Spin-polarized
  states, spin zero modes, and quantum hall effect.
\newblock {\em Phys. Rev. B}, 79:161409, Apr 2009.

\bibitem{BR_Graphene_Offidani_17}
Manuel Offidani, Mirco Milletar\`{\i}, Roberto Raimondi, and Aires Ferreira.
\newblock Optimal charge-to-spin conversion in graphene on transition-metal
  dichalcogenides.
\newblock {\em Phys. Rev. Lett.}, 119:196801, Nov 2017.

\bibitem{Ghiasi_19}
Talieh~S. Ghiasi, Alexey~A. Kaverzin, Patrick~J. Blah, and Bart~J. van Wees.
\newblock Charge-to-spin conversion by the rashba--edelstein effect in
  two-dimensional van der waals heterostructures up to room temperature.
\newblock {\em Nano Letters}, 19(9):5959--5966, 09 2019.

\bibitem{Benitez_20}
L.~Antonio Ben{\'\i}tez, Williams Savero~Torres, Juan~F. Sierra, Matias
  Timmermans, Jose~H. Garcia, Stephan Roche, Marius~V. Costache, and Sergio~O.
  Valenzuela.
\newblock Tunable room-temperature spin galvanic and spin hall effects in van
  der waals heterostructures.
\newblock {\em Nature Materials}, 19(2):170--175, 2020.

\bibitem{Li_2020}
Lijun Li, Jin Zhang, Gyuho Myeong, Wongil Shin, Hongsik Lim, Boram Kim, Seungho
  Kim, Taehyeok Jin, Stuart Cavill, Beom~Seo Kim, Changyoung Kim, Johannes
  Lischner, Aires Ferreira, and Sungjae Cho.
\newblock Gate-tunable reversible rashba--edelstein effect in a few-layer
  graphene/2h-tas2 heterostructure at room temperature.
\newblock {\em ACS Nano}, 14(5):5251--5259, 05 2020.

\bibitem{Hoque_21}
Anamul~Md. Hoque, Dmitrii Khokhriakov, Klaus Zollner, Bing Zhao, Bogdan
  Karpiak, Jaroslav Fabian, and Saroj~P. Dash.
\newblock All-electrical creation and control of spin-galvanic signal in
  graphene and molybdenum ditelluride heterostructures at room temperature.
\newblock {\em Communications Physics}, 4(1):124, 2021.

\bibitem{Kondou_16}
K.~Kondou, R.~Yoshimi, A.~Tsukazaki, Y.~Fukuma, J.~Matsuno, K.~S. Takahashi,
  M.~Kawasaki, Y.~Tokura, and Y.~Otani.
\newblock Fermi-level-dependent charge-to-spin current conversion by dirac
  surface states of topological insulators.
\newblock {\em Nature Physics}, 12(11):1027--1031, 2016.

\bibitem{Ghiasi2017}
Talieh~S. Ghiasi, Josep Ingla-Ayn{\'{e}}s, Alexey~A. Kaverzin, and Bart~J. van
  Wees.
\newblock Large proximity-induced spin lifetime anisotropy in transition-metal
  dichalcogenide/graphene heterostructures.
\newblock {\em Nano Letters}, 17(12):7528--7532, November 2017.

\bibitem{Benitez2017}
L.~Antonio Ben{\'{\i}}tez, Juan~F. Sierra, Williams~Savero Torres, Aloïs
  Arrighi, Fr{\'{e}}d{\'{e}}ric Bonell, Marius~V. Costache, and Sergio~O.
  Valenzuela.
\newblock Strongly anisotropic spin relaxation in
  graphene{\textendash}transition metal dichalcogenide heterostructures at room
  temperature.
\newblock {\em Nature Physics}, 14(3):303--308, December 2017.

\bibitem{Cummings2017}
Aron~W. Cummings, Jose~H. Garcia, Jaroslav Fabian, and Stephan Roche.
\newblock Giant spin lifetime anisotropy in graphene induced by proximity
  effects.
\newblock {\em Phys. Rev. Lett.}, 119:206601, Nov 2017.

\bibitem{Manuel2018}
Manuel Offidani and Aires Ferreira.
\newblock Microscopic theory of spin relaxation anisotropy in graphene with
  proximity-induced spin-orbit coupling.
\newblock {\em Phys. Rev. B}, 98:245408, Dec 2018.

\bibitem{BR_Graphene_Milletari_17}
Mirco Milletar\`{\i}, Manuel Offidani, Aires Ferreira, and Roberto Raimondi.
\newblock Covariant conservation laws and the spin hall effect in dirac-rashba
  systems.
\newblock {\em Phys. Rev. Lett.}, 119:246801, Dec 2017.

\bibitem{BR_Graphene_Offidani_MDPI_18}
Manuel Offidani, Roberto Raimondi, and Aires Ferreira.
\newblock Microscopic linear response theory of spin relaxation and
  relativistic transport phenomena in graphene.
\newblock {\em Condensed Matter}, 3:18, 2018.

\bibitem{Offidani_Ferreira_18}
Manuel Offidani and Aires Ferreira.
\newblock Anomalous hall effect in 2d dirac materials.
\newblock {\em Phys. Rev. Lett.}, 121:126802, Sep 2018.

\bibitem{Sousa_2020}
Frederico Sousa, Gen Tatara, and Aires Ferreira.
\newblock Skew-scattering-induced giant antidamping spin-orbit torques:
  Collinear and out-of-plane edelstein effects at two-dimensional
  material/ferromagnet interfaces.
\newblock {\em Phys. Rev. Research}, 2:043401, Dec 2020.

\bibitem{Cavill_20}
Stuart~A. Cavill, Chunli Huang, Manuel Offidani, Yu-Hsuan Lin, Miguel~A.
  Cazalilla, and Aires Ferreira.
\newblock Proposal for unambiguous electrical detection of spin-charge
  conversion in lateral spin valves.
\newblock {\em Phys. Rev. Lett.}, 124:236803, Jun 2020.

\bibitem{Hirsch_99}
J.~E. Hirsch.
\newblock Spin hall effect.
\newblock {\em Phys. Rev. Lett.}, 83:1834--1837, Aug 1999.

\bibitem{Garcia_18}
Jose~H. Garcia, Marc Vila, Aron~W. Cummings, and Stephan Roche.
\newblock Spin transport in graphene/transition metal dichalcogenide
  heterostructures.
\newblock {\em Chem. Soc. Rev.}, 47:3359--3379, 2018.

\bibitem{Joao_20}
Simao~M. Joao, Misa Andelkovic, Lucian Covaci, Tatiana~G. Rappoport, Joao M.
  V.~P. Lopes, and Aires Ferreira.
\newblock Kite: high-performance accurate modelling of electronic structure and
  response functions of large molecules, disordered crystals and
  heterostructures.
\newblock {\em Royal Society Open Science}, 7(2):191809, 2020.

\bibitem{Tuan_16}
D.~Van~Tuan, J.~M. Marmolejo-Tejada, X.~Waintal, B.~K.
  Nikoli\ifmmode~\acute{c}\else \'{c}\fi{}, S.~O. Valenzuela, and S.~Roche.
\newblock Spin hall effect and origins of nonlocal resistance in
  adatom-decorated graphene.
\newblock {\em Phys. Rev. Lett.}, 117:176602, Oct 2016.

\bibitem{review_valleytronics_16}
John~R. Schaibley, Hongyi Yu, Genevieve Clark, Pasqual Rivera, Jason~S. Ross,
  Kyle~L. Seyler, Wang Yao, and Xiaodong Xu.
\newblock Valleytronics in 2d materials.
\newblock {\em Nature Reviews Materials}, 1(11):16055, 2016.

\bibitem{Xiao_12}
Di~Xiao, Gui-Bin Liu, Wanxiang Feng, Xiaodong Xu, and Wang Yao.
\newblock Coupled spin and valley physics in monolayers of ${\mathrm{mos}}_{2}$
  and other group-vi dichalcogenides.
\newblock {\em Phys. Rev. Lett.}, 108:196802, May 2012.

\bibitem{Luo_17}
Yunqiu~Kelly Luo, Jinsong Xu, Tiancong Zhu, Guanzhong Wu, Elizabeth~J.
  McCormick, Wenbo Zhan, Mahesh~R. Neupane, and Roland~K. Kawakami.
\newblock Opto-valleytronic spin injection in monolayer mos2/few-layer graphene
  hybrid spin valves.
\newblock {\em Nano Letters}, 17(6):3877--3883, 06 2017.

\bibitem{Avsar_17}
Ahmet Avsar, Dmitrii Unuchek, Jiawei Liu, Oriol~Lopez Sanchez, Kenji Watanabe,
  Takashi Taniguchi, Barbaros {\"O}zyilmaz, and Andras Kis.
\newblock Optospintronics in graphene via proximity coupling.
\newblock {\em ACS Nano}, 11(11):11678--11686, 11 2017.

\bibitem{Wang_11}
P.~Zhang and M.~W. Wu.
\newblock Electron spin diffusion and transport in graphene.
\newblock {\em Phys. Rev. B}, 84:045304, Jul 2011.

\bibitem{2DEG_SOC_Burkov_04}
A.~A. Burkov, Alvaro~S. N\'u\~nez, and A.~H. MacDonald.
\newblock Theory of spin-charge-coupled transport in a two-dimensional electron
  gas with rashba spin-orbit interactions.
\newblock {\em Phys. Rev. B}, 70:155308, Oct 2004.

\bibitem{Wenk_10}
P.~Wenk and S.~Kettemann.
\newblock Dimensional dependence of weak localization corrections and spin
  relaxation in quantum wires with rashba spin-orbit coupling.
\newblock {\em Phys. Rev. B}, 81:125309, Mar 2010.

\bibitem{BR_Bychkov_Rashba_84}
Y.~A. Bychkov and E.~I. Rashba.
\newblock Properties of a 2d electron gas with lifted spectral degeneracy.
\newblock {\em JETP Letters}, 39(2), 1984.

\bibitem{note:1}
Within the single-valley representation adopted in this work, the Pauli
  matrices $\sigma_{a}$ and $s_{b}$ all anticommute with $\mathcal{T}$, so that
  their products are invariant under the time-reversal operation. The
  Hamiltonian at the $K^{\prime}$ point is obtained by simply changing the sign
  of the time-like gauge-field component
  $\mathcal{A}_{K}^{0}=-\mathcal{A}_{K^{\prime}}^{0}=\lambda_{\textrm{sv}}s_{z}$,
  known in the literature as valley-Zeeman interaction or spin-valley coupling.

\bibitem{Pachoud_14}
Alexandre Pachoud, Aires Ferreira, B.~\"Ozyilmaz, and A.~H. Castro~Neto.
\newblock Scattering theory of spin-orbit active adatoms on graphene.
\newblock {\em Phys. Rev. B}, 90:035444, Jul 2014.

\bibitem{Yang_16}
H.-Y. Yang, Chunli Huang, H.~Ochoa, and M.~A. Cazalilla.
\newblock Extrinsic spin hall effect from anisotropic rashba spin-orbit
  coupling in graphene.
\newblock {\em Phys. Rev. B}, 93:085418, Feb 2016.

\bibitem{Huang_16}
Chunli Huang, Y.~D. Chong, and Miguel~A. Cazalilla.
\newblock Direct coupling between charge current and spin polarization by
  extrinsic mechanisms in graphene.
\newblock {\em Phys. Rev. B}, 94:085414, Aug 2016.

\bibitem{SOC_Graphene_Milletari_Ferreira16a}
Mirco Milletar\`{\i} and Aires Ferreira.
\newblock Quantum diagrammatic theory of the extrinsic spin hall effect in
  graphene.
\newblock {\em Phys. Rev. B}, 94:134202, Oct 2016.

\bibitem{Milletari_16_b}
Mirco Milletar\`{\i} and Aires Ferreira.
\newblock Crossover to the anomalous quantum regime in the extrinsic spin hall
  effect of graphene.
\newblock {\em Phys. Rev. B}, 94:201402, Nov 2016.

\bibitem{RevModPhys.76.323}
Igor \ifmmode \check{Z}\else \v{Z}\fi{}uti\ifmmode~\acute{c}\else \'{c}\fi{},
  Jaroslav Fabian, and S.~Das~Sarma.
\newblock Spintronics: Fundamentals and applications.
\newblock {\em Rev. Mod. Phys.}, 76:323--410, Apr 2004.

\bibitem{Wu_review}
M.W. Wu, J.H. Jiang, and M.Q. Weng.
\newblock Spin dynamics in semiconductors.
\newblock {\em Physics Reports}, 493(2):61--236, 2010.

\bibitem{Boross_13}
P{\'e}ter Boross, Bal{\'a}zs D{\'o}ra, Annam{\'a}ria Kiss, and Ferenc Simon.
\newblock A unified theory of spin-relaxation due to spin-orbit coupling in
  metals and semiconductors.
\newblock {\em Scientific Reports}, 3(1):3233, 2013.

\bibitem{Raes_16}
Bart Raes, Jeroen~E. Scheerder, Marius~V. Costache, Fr{\'e}d{\'e}ric Bonell,
  Juan~F. Sierra, Jo~Cuppens, Joris Van~de Vondel, and Sergio~O. Valenzuela.
\newblock Determination of the spin-lifetime anisotropy in graphene using
  oblique spin precession.
\newblock {\em Nature Communications}, 7(1):11444, 2016.

\bibitem{Ringer_18}
Sebastian Ringer, Stefan Hartl, Matthias Rosenauer, Tobias V\"olkl, Maximilian
  Kadur, Franz Hopperdietzel, Dieter Weiss, and Jonathan Eroms.
\newblock Measuring anisotropic spin relaxation in graphene.
\newblock {\em Phys. Rev. B}, 97:205439, May 2018.

\bibitem{Book_Fetter}
A.~L. Fetter and J.~D. Walecka.
\newblock {\em Quantum Theory of Many-Particle Systems}.
\newblock McGraw-Hill, Boston, 1971.

\bibitem{Doniach}
S~Doniach and E~H Sondheimer.
\newblock {\em Green's Functions for Solid State Physicists}.
\newblock Imperial College Press, London, 1998.

\bibitem{Mahan}
G.~D. Mahan.
\newblock {\em Many Particle Physics, Third Edition}.
\newblock Plenum, New York, 2000.

\bibitem{Ado_2015}
I.~A. Ado, I.~A. Dmitriev, P.~M. Ostrovsky, and M.~Titov.
\newblock Anomalous hall effect with massive dirac fermions.
\newblock {\em {EPL} (Europhysics Letters)}, 111(3):37004, aug 2015.

\bibitem{note:2}
This is markdly different in models with noncoplanar spin texture, e.g. caused
  by the interplay of BR effect and spin-valley coupling. The resulting net
  $\hat{z}$-spin polarization at each valley activates a robust skew scattering
  mechanism, whose description relies on high-order scattering processes that
  are best described within the T-matrix framework
  \cite{BR_Graphene_Milletari_17,BR_Graphene_Offidani_MDPI_18}. A detailed
  discussion on the validity of the Gaussian approximation for 2D Dirac
  fermions is given in Ref. \cite{SOC_Graphene_Milletari_Ferreira16a}.

\bibitem{Shen_14}
Ka~Shen, R.~Raimondi, and G.~Vignale.
\newblock Theory of coupled spin-charge transport due to spin-orbit interaction
  in inhomogeneous two-dimensional electron liquids.
\newblock {\em Phys. Rev. B}, 90:245302, Dec 2014.

\bibitem{Burkov_10}
A.~A. Burkov and D.~G. Hawthorn.
\newblock Spin and charge transport on the surface of a topological insulator.
\newblock {\em Phys. Rev. Lett.}, 105:066802, Aug 2010.

\bibitem{Sacksteder_12}
Vincent~E. Sacksteder, Stefan Kettemann, QuanSheng Wu, Xi~Dai, and Zhong Fang.
\newblock Spin conduction in anisotropic three-dimensional topological
  insulators.
\newblock {\em Phys. Rev. B}, 85:205303, May 2012.

\bibitem{Basko_08}
D.~M. Basko.
\newblock Theory of resonant multiphonon raman scattering in graphene.
\newblock {\em Phys. Rev. B}, 78:125418, Sep 2008.

\bibitem{RevModPhys.89.011001}
John Schliemann.
\newblock Colloquium: Persistent spin textures in semiconductor nanostructures.
\newblock {\em Rev. Mod. Phys.}, 89:011001, Jan 2017.

\bibitem{Cysne_18}
Tarik~P. Cysne, Aires Ferreira, and Tatiana~G. Rappoport.
\newblock Crystal-field effects in graphene with interface-induced spin-orbit
  coupling.
\newblock {\em Phys. Rev. B}, 98:045407, Jul 2018.

\bibitem{SOC_2DEG_Ka_Raimondi_Vignale_14b}
Ka~Shen, G.~Vignale, and R.~Raimondi.
\newblock Microscopic theory of the inverse edelstein effect.
\newblock {\em Phys. Rev. Lett.}, 112:096601, Mar 2014.

\bibitem{note:3}
In the absence of intervalley scattering channels (i.e., for scalar
  impurities), the transport time of 2D Dirac fermions is twice the scattering
  time $\tau_{\textrm{tr}}=2\tau$ due to the supression of backscattering
  events.

\bibitem{Kato1910}
Y.~K. Kato, R.~C. Myers, A.~C. Gossard, and D.~D. Awschalom.
\newblock Observation of the spin hall effect in semiconductors.
\newblock {\em Science}, 306(5703):1910--1913, 2004.

\bibitem{Wunderlich_05}
J.~Wunderlich, B.~Kaestner, J.~Sinova, and T.~Jungwirth.
\newblock Experimental observation of the spin-hall effect in a two-dimensional
  spin-orbit coupled semiconductor system.
\newblock {\em Phys. Rev. Lett.}, 94:047204, Feb 2005.

\bibitem{Dimitrova_05}
Ol'ga~V. Dimitrova.
\newblock Spin-hall conductivity in a two-dimensional rashba electron gas.
\newblock {\em Phys. Rev. B}, 71:245327, Jun 2005.

\bibitem{Aranov_89}
A.~G. Aranov and Y.~B. Lyanda-Geller.
\newblock Spin polarization of electrons by an electric current.
\newblock {\em JETP Lett.}, 50:431, 1989.

\bibitem{BR_Edelstein90}
V.M. Edelstein.
\newblock Spin polarization of conduction electrons induced by electric current
  in two-dimensional asymmetric electron systems.
\newblock {\em Solid State Communications}, 73(3):233 -- 235, 1990.

\bibitem{Bludov_13}
Y.~V. Bludov, Aires Ferreira, N.~M.~R. Peres, and M.~I. Vasilevskiy.
\newblock A primer on surface plasmon-polaritons in graphene.
\newblock {\em International Journal of Modern Physics B}, 27(10):1341001,
  2013.

\bibitem{SOC_2DEG_Ka_Raimondi_Vignale_14a}
Ka~Shen, R.~Raimondi, and G.~Vignale.
\newblock Theory of coupled spin-charge transport due to spin-orbit interaction
  in inhomogeneous two-dimensional electron liquids.
\newblock {\em Phys. Rev. B}, 90:245302, Dec 2014.

\end{thebibliography}

\end{document}